\documentclass[floatfix,aps,showpacs]{revtex4}
\usepackage{pstricks}
\usepackage{pst-coil}
\usepackage{amssymb}
\usepackage{graphics,epsf}
\usepackage{amsmath}

\numberwithin{equation}{section}
\renewcommand{\theequation}{\arabic{section}.\arabic{equation}}

\newcommand{\be}{\begin{eqnarray}}
\newcommand{\ee}{\end{eqnarray}}
\newcommand{\bes}{\begin{eqnarray*}}
\newcommand{\ees}{\end{eqnarray*}}
\newcommand{\bmt}[1]{\mbox{\boldmath $#1$}}
\newcommand{\f}{\frac}

\newcommand{\p}{\partial}
\newcommand{\w}{\tilde}

\newcommand{\ov}{\overline}
\newcommand{\wh}{\widehat}
\newcommand{\ph}{\phantom}
\newcommand{\cp}{{\cal{P}}}
\newtheorem{theorem}{Theorem}[section]
\newtheorem{lemma}{Lemma}[section]
\newtheorem{corollary}{Corollary}[section]

\begin{document}

\title{Even perturbations of self-similar Vaidya space-time.}
\author{Brien C Nolan\footnote{Electronic address:
brien.nolan@dcu.ie} and Thomas J Waters\footnote{Electronic
address: thomas.waters2@mail.dcu.ie}} \affiliation{School of
Mathematical Sciences, Dublin City University, Glasnevin, Dublin
9, Ireland.}
\begin{abstract}
We study even parity metric and matter perturbations of all
angular modes in self-similar Vaidya space-time. We focus on the
case where the background contains a naked singularity. Initial
conditions are imposed describing a finite perturbation emerging
from the portion of flat space-time preceding the matter-filled
region of space-time. The most general perturbation satisfying the
initial conditions is allowed impinge upon the Cauchy horizon
(CH), whereat the perturbation remains finite: there is no
``blue-sheet'' instability. However when the perturbation evolves
through the CH and onto the second future similarity horizon of
the naked singularity, divergence necessarily occurs: this surface
is found to be unstable. The analysis is based on the study of
individual modes following a Mellin transform of the perturbation.
We present an argument that the full perturbation remains finite
after resummation of the (possibly infinite number of) modes.
\end{abstract}
 \maketitle

\section{Introduction and summary.}

The Cosmic Censorship Hypothesis (CCH) was proposed by Penrose and
states that naked singularities will not evolve from generic
initial data (see Wald \cite{wald} for detailed discussion). The
motivation is obvious: information leaving the singularity is
impossible to predict and all physical laws in the region affected
by the naked singularity will break down. While the CCH seems
intuitively reasonable, there are some notable counter-examples,
for instance: the Reissner-Nordstr\"om and Kerr timelike
singularities, certain classes of self-similar perfect fluid
\cite{OP} and dust \cite{Joshi} solutions, and the self-similar
scalar field \cite{Christ}. In general, self-similar spherically
symmetric (SSSS) space-times are a rich source of counter-examples
to the CCH.

These naked singularities occur in space-times with very high
degrees of symmetry, and thus perhaps the naked singularities are
due to these unphysical restraints. We seek to move away from this
symmetry by perturbing the space-time and examining if the
singularity remains naked; that is we look for stability.

In the Reissner-Nordstr\"om space-time, Penrose \cite{Penrose}
noticed that the Cauchy horizon (CH), that surface marking the
boundary between regions where the singularity is and is not
visible to observers, would be a surface of infinite blue shift.
Building on this, Chandrasekhar and Hartle \cite{chandhartle}
found that metric perturbations originating from the exterior have
a flux that observers crossing the CH will measure as infinite.
Thus the CH has a ``blue sheet'' instability, which prevents
observers from crossing the CH and observing the singularity. This
result has been firmly established in further work (see
\cite{brady1} and \cite{Dafermos}). This suggests the Cauchy
horizon is perhaps unstable in other possible counter-examples to
Cosmic Censorship. We will examine this mechanism of instability
in SSSS space-times.

In a previous paper \cite{bandt}, the authors have shown that all
self-similar spherically symmetric space-times which admit a
Cauchy horizon, and thus a naked singularity, are stable when we
inject a scalar field; that is, the flux of the scalar field as
measured by a timelike observer crossing the Cauchy horizon
remains finite. This paper extends on that work by considering
metric and matter perturbations of SSSS space-times. In related
work, Miyamoto and Harada \cite{HaMi} showed that the CH in the
general class of space-times mentioned above is unstable at the
semi-classical level.

For the perturbations we will use the formalism of Gerlach and
Sengupta \cite{GS}. They describe gauge invariant metric and
matter perturbations of spherically symmetric space-times by a 2+2
split and multipole decomposition. This formalism has been used
for a variety of backgrounds, for example Schwarzschild
\cite{Sarbach}, timelike dust \cite{TH} and perfect fluid
\cite{Gund}. As we must specify the matter content we have begun
by considering the self-similar spherically symmetric null dust
(i.e.\ self-similar Vaidya) space-time. This space-time is of
interest because it provides a mathematically simple (self-similar
and spherically symmetric) example of an exact solution of
Einstein's equation (that is, the metric can be written explicitly
in terms of elementary functions) with an energy-momentum tensor
obeying the dominant energy condition that admits a naked
singularity for an easily specified set of parameters. Waugh and
Lake \cite{WL1} showed the stability of the Cauchy horizon in
Vaidya space-time at the level of the eikonal approximation.

In the following section we define the background space-time and
give the conditions for a naked singularity. We also introduce the
Mellin transform and point out how the partial differential
perturbation equations can be expressed as parametrised ordinary
differential equations: the perturbation is then given by the
inverse Mellin transform (resummation) of the solutions of the
ODEs. The analysis of Sections 3 - 5 deals with these ODEs and
their solutions (the `modes'). In section 3, we describe the
gauge-invariant formalism of Gerlach and Sengupta. While metric
perturbations are primarily useful in modelling gravitational
radiation which manifests in quadrupolar moments and above (see,
for example, \cite{MTW}), for completeness we will consider all
multipole moments of the perturbation.

In section 4 we begin specifying our initial data. For
completeness we must take into account perturbations arising from
the region of flat space preceding the matter-filled Vaidya
space-time. We will denote the boundary between these two regions
$\mathcal{N}$, and define $\mathcal{N}$ as $x=0$ where $x<0$ is
flat space-time and $x>0$ is Vaidya space-time; here $x$ is a
naturally occurring similarity variable. Our initial/boundary
conditions are: i) finiteness of the perturbations on the axis
$r=0$ in the Minkowski region, ii) finiteness on $\mathcal{N}$ as
$x\uparrow 0$, iii) finiteness on $\mathcal{N}$ as $x\downarrow
0$, and iv) continuous matching across $x=0$.

When we solve the perturbation equations for multipole mode number
$l\geq 2$, we find a two-parameter family of solutions in the
portion of Minkowski space-time. To satisfy initial condition (i),
we must set one of these parameters to zero. This specifies the
type of solutions which will solve initial condition (ii). The
second initial condition restricts the acceptable range of the
mode number $s$ that arises from the Mellin transform of the
perturbation equations; that is we can only consider $s \geq 2$,
$s \neq l+m$ where $m\in \mathbb{N}$. In section 5 we solve the
perturbation equations in the Vaidya region, and we find a four
parameter family of solutions. However, to solve initial condition
(iii), we must further restrict the range of $s$ to $s >2$, $s
\notin \mathbb{Z}$. To satisfy initial condition (iv), one of the
parameters is fixed and thus we have a three parameter family of
solutions.

This class of perturbations satisfying the initial conditions is
then allowed evolve onto the Cauchy horizon, and there it is found
that the flux of the perturbation is finite. Thus the Cauchy
horizon is stable under metric and matter perturbations ($l \geq
2$), and does not display the blue-sheet instability shown by
Reissner-Nordstr\"om space-time. Interestingly, when the
perturbations are allowed evolve through the Cauchy horizon and
onto the second future similarity horizon (see below), here the
flux of the perturbations will diverge, meaning that this horizon
is \emph{unstable} under metric and matter perturbations ($l\geq
2$).

For the dipole mode $l=1$, we reproduce the familiar result that
the perturbation in the Minkowski region is pure gauge, that is we
can always find a coordinate system in which the $l=1$
perturbation vanishes. In the Vaidya region there is an $l=1$
perturbation which cannot be gauged away, and this evolves through
the Vaidya space-time without divergence.

The $l=0$ mode is spherically symmetric, and we can use uniqueness
results: in the Minkowski space-time the $l=0$ perturbation
generates a Schwarzschild space-time, which has to have its mass
set to zero to satisfy initial condition 1. In the Vaidya region,
after a spherically symmetric perturbation we again have Vaidya
space-time, merely with a new mass function and null coordinate.

In Section 6, we consider the problem of resummation: does
finiteness of the individual modes for an allowed range of the
parameter $s$ imply finiteness of the full perturbation given by
the inverse Mellin transform over $s$ of the modes? We present
what we consider a plausibility argument that the answer to this
question is `yes': a complete rigorous proof is beyond the scope
of the present paper. We conclude with a brief discussion in
Section 7. We use the conventions of Wald \cite{wald} and set
$c=G=1$.

\section{Self-similar Vaidya space-time.}

The matter model we consider is the in-falling null dust or Vaidya
model. Its metric is found from the Schwarzschild metric by
replacing the constant mass term $m$ with a function of the
advanced (ingoing) Bondi coordinate $v$: \be g_{\mu\nu}dx^\mu
dx^\nu=-\left(1-\f{m(v)}{r}\right)dv^2+2dv dr+r^2 d\Omega^2, \ee
where $d\Omega^2$ is the line element of the unit 2-sphere. By
setting $m$ linear, that is $m=\lambda v$, the space-time becomes
self-similar; that is there is a homothetic Killing vector field
\be \vec{\zeta}=v \f{\p}{\p v}+r\f{\p}{\p r}, \ee such that
$\mathcal{L}_{\vec{\zeta}}g_{\mu\nu}=2 g_{\mu\nu}$. Thus we can
introduce the similarity coordinate $x=v/r$, and in $(x,r)$
coordinates the self-similar Vaidya metric and matter tensors
become \be g_{\mu\nu}dx^\mu dx^\nu &=& r^2\left(-1+\lambda
x\right)dx^2+2r\left(1 - x + \lambda x^2\right)dx
dr+x\left(2-x+\lambda x^2\right)dr^2+r^2 d\Omega^2,
\label{Vmetric} \\ t_{\mu\nu}dx^\mu dx^\nu &=& \rho \ell_\mu
\ell_\nu dx^\mu dx^\nu= \f{\lambda}{8 \pi } dx^2+\f{\lambda x}{8
\pi r } dx dr+\f{\lambda x^2}{8 \pi r^2} dr^2,
 \label{Vmatter}\ee where $\rho=\dot{m}(v)/8 \pi r^2=\lambda/8 \pi  r^2$ is
the energy density,
$\ell_\mu=-\p_\mu v$
is the ingoing null direction, and Minkowski space-time is
recovered in the limit $\lambda\rightarrow 0$.

\begin{figure}
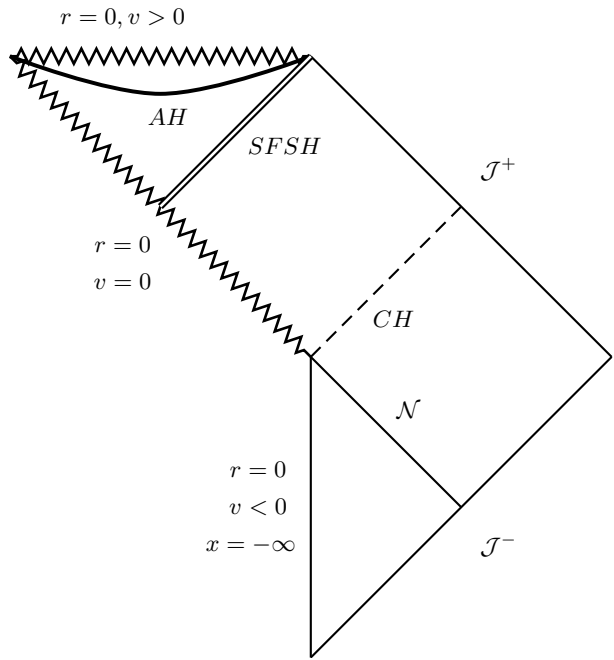

\pspicture*(0,0)(9,9) $ \psline(4,0)(4,4) \psline(4,0)(8,4)
\psline(8,4)(4,8) \psline(6,2)(4,4)
\psline[linestyle=dashed](4,4)(6,6)
\pszigzag[coilwidth=.2,coilarm=.1,linewidth=1pt](4,4)(0,8)
\pszigzag[coilwidth=.2,coilarm=.1,linewidth=1pt](0,8)(4,8)
\pscurve[linewidth=1.5pt](0,8)(2,7.5)(4,8)
\psline[doubleline=true,doublesep=1pt](2,6)(4,8)
\rput(6.5,1.5){\psframebox*[framearc=.3]{{\cal J}^-}}
\rput(6.5,6.5){\psframebox*[framearc=.3]{{\cal J}^+}}
\rput(3.3,2.5){\psframebox*[framearc=.3]{r=0}}
\rput(3.3,2){\psframebox*[framearc=.3]{v<0}}
\rput(3.2,1.5){\psframebox*[framearc=.3]{x=-\infty}}
\rput(1.5,8.5){\psframebox*[framearc=.3]{r=0,v>0}}
\rput(1.5,5.5){\psframebox*[framearc=.3]{r=0}}
\rput(1.5,5){\psframebox*[framearc=.3]{v=0}}
\rput(5.1,4.5){\psframebox*[framearc=.3]{CH}}
\rput(3.65,6.8){\psframebox*[framearc=.3]{SFSH}}
\rput(2.1,7.2){\psframebox*[framearc=.3]{AH}}
\rput(5.3,3.3){\psframebox*[framearc=.3]{{\cal{N}}}}  $
\endpspicture
\caption{Conformal diagram for Vaidya space-time admitting a
globally naked singularity. The region preceding $\mathcal{N}$ is
a portion of Minkowski space-time, and the region to the future of
$\mathcal{N}$ is filled with null dust. There are three similarity
horizons at which the similarity coordinate $x$ is null: $x=0$
denoted $\cal{N}$, $x=x_c$ shown dashed, and $x=x_e$ shown as a
double line. We identify $x=x_c$ as the Cauchy horizon, and will
call $x=x_e$ the second future similarity horizon (SFSH). The
apparent horizon given by $x=1/\lambda$ is shown as a bold curve.}
\end{figure}

The matter field is switched on at $v=x=0$ (which we will call the
``threshold'' and is denoted $\cal{N}$ in Figure 1), the region
$v<0$ being flat. When the matter collapses to the origin of
coordinates a singularity forms. We can describe the global
structure of the space-time by analysing the causal nature of the
similarity coordinate $x$. Null homothetic lines are zeroes of
$g_{rr}$ given in (\ref{Vmetric}), that is $x(\lambda x^2-x+2)$,
thus $x=0$ is the null homothetic line pointing radially inward to
the singularity at the scaling origin (its past null cone,
$\mathcal{N}$). If there are other positive real zeroes of
$\lambda x^2-x+2$, then these represent null homothetic lines
emanating to future null infinity which meet the singularity in
the past. The lowest of these zeroes is thus the first null
geodesic to leave the singularity and escape to future null
infinity: the Cauchy horizon. It is given by \be
x=\f{1}{2\lambda}(1-\sqrt{1-8\lambda})\equiv x_c, \ee and exists
for $0<\lambda <1/8$. Subsequent zeroes are additional
``similarity horizons'': they mark the transition of $x$ from
timelike to spacelike or vice versa. For the self-similar Vaidya
space-time, there is one more similarity horizon, \be
x=\f{1}{2\lambda}(1+\sqrt{1-8\lambda})\equiv x_e. \ee For
$0<\lambda <1/8$ these similarity horizons are distinct and the
singularity is globally naked. In the notation of Carr and
Gundlach \cite{CarrGund} the causal structure is rTfSfTpSs, and is
shown in Figure 1. For $\lambda=1/8$ these horizons coincide and
the singularity is instantaneously (marginally) naked; we will not
consider this case in this paper. For $\lambda > 1/8$ a black hole
forms, see Figure 2.

The second similarity horizon is, in the purely self-similar
Vaidya case, the last null geodesic to leave the singularity and
escape to future null infinity, and thus can be called the event
horizon. However, to have an asymptotically flat model, we can
match across $v=v_+>0$ with the exterior Schwarzschild space-time;
in this case $x=x_e$ would not be the event horizon. Thus we will
call $x=x_e$ the second future similarity horizon (SFSH).

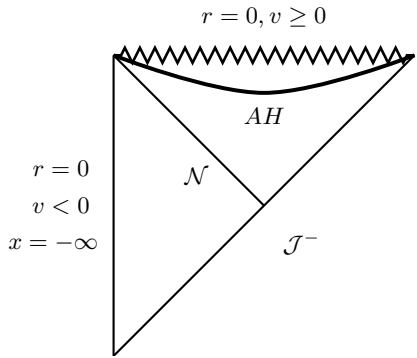
\begin{figure}
\begin{pspicture*}(0,0)(9,6) $ \psline(3,0)(3,4) \psline(3,0)(7,4)
\pszigzag[coilwidth=.2,coilarm=.1,linewidth=1pt](3,4)(7,4)
\psline(5,2)(3,4) \rput(2.3,2.5){\psframebox*[framearc=.3]{r=0}}
\rput(2.3,2){\psframebox*[framearc=.3]{v<0}}
\rput(2.2,1.5){\psframebox*[framearc=.3]{x=-\infty}}
\rput(5.5,1.5){\psframebox*[framearc=.3]{{\cal J}^-}}
\pscurve[linewidth=1.5pt](3,4)(5,3.5)(7,4)
\rput(5,3.2){\psframebox*[framearc=.3]{AH}}
\rput(4.1,2.4){\psframebox*[framearc=.3]{{\cal{N}}}}
\rput(5,4.5){\psframebox*[framearc=.3]{r=0,v\geq 0}} $
\end{pspicture*}
\caption{Conformal diagram for Vaidya space-time with censored
singularity, corresponding to the case $\lambda >1/8$. A spacelike
singularity forms at $r=0$ for $v\geq0$.}
\end{figure}

Our use of the coordinates $(x,r)$ is motivated by the fact that
they are global, important surfaces are easily defined in terms of
the $x$-coordinate, and the entire space-time (Minkowski and
Vaidya) is covered by $r\in [0,\infty)$. Then using the
Mellin-transform the globally defined partial differential
perturbation equations reduce to ordinary differential equations;
this occurs precisely because our space-time is self-similar. The
Mellin transform is defined by \be G(x;s)={\cal{M}}[g(x,r)](r\to
s):=\int_0^{\infty} g(x,r) r^{s-1} dr, \label{mell}\ee with
$s\in\mathbb{C}$. In coordinates $(x,r)$, partial derivatives with
respect to $r$ only appear in the perturbation equations in the
form $r\frac{\partial}{\partial r}$ and so the Mellin transform of
these equations involves only derivatives with respect to $x$ and
coefficients independent of $r$. This amounts to replacing
$g(x,r)$ with $r^s G(x;s)$, in the same way that Laplace
transforms amount to replacing $g(x,r)$ with $e^{sr} G(x;s)$. The
solution is recovered by performing an inverse Mellin transform,
i.e.\ by integrating over the valid range of $s$. See Section 6
for a discussion of the validity of this procedure and details on
the Mellin transform. The following analysis is carried out for
what we will refer to as the `modes', i.e.\ the parametrised (by
complex $s$) functions $G(x;s)$.

\section{Perturbation Formalism.}

Next we review the formalism of Gerlach and Sengupta \cite{GS}. We
exploit the spherical symmetry of the background by performing a
2+2 split of the space-time into $M^2$ (manifold spanned by time
and radial coordinates) and $S^2$ (unit 2-sphere), and then
decomposing the angular part of the
perturbation in terms of spherical harmonics.\\
First some notation: Greek indices represent coordinates on the
four-dimensional space-time, capital Latin indices coordinates on
$M^2$, lower case Latin indices $S^2$, and we define covariant
derivatives so: \be g_{\mu\nu;\lambda}=0, \quad g_{AB|C}=0, \quad
g_{ab:c}=0, \ee and a comma defines an ordinary partial
derivative.

\subsection{Angular decomposition.}

We write a non-spherical metric perturbation \be g_{\mu
\nu}=\w{g}_{\mu \nu}+h_{\mu \nu}(t,r,\theta,\phi), \ee where the
over-tilde denotes background quantities; and similarly for the
matter perturbation \be t_{\mu\nu}=\w{t}_{\mu\nu}+\Delta
t_{\mu\nu}(t,r,\theta,\phi). \ee From the spherical harmonics we
can construct bases for vectors, \be Y,_a \quad ; \quad
S_a=\epsilon_a^{\phantom{a} b} Y,_b \label{vecbases} \ee and
tensors, \be Y\gamma_{ab} \quad ; \quad
Z_{ab}=Y,_{a:b}+\f{1}{2}l(l+1)Y\gamma_{ab} \quad ; \quad
S_{(a:b)}, \label{tenbases} \ee where summation over the mode
numbers $l,m$ is implied. Using these, we decompose the
perturbation in terms of scalar, vector and tensor objects defined
on $M^2$, times scalar, vector and tensor bases defined on $S^2$.

When we compute the linearized Einstein equations for a perturbed
space-time decomposed in this way, we find they naturally decouple
into two sectors, even and odd. Gundlach and
Mart\'{\i}n-Garc\'{\i}a's \cite{Gund} definition is the most
straightforward: that sector whose bases are in even powers of
$\epsilon_a^{\phantom{a} b}$ is called even (or polar or
spheroidal); that sector whose bases are in odd powers of
$\epsilon_a^{\phantom{a} b}$ is called odd (or axial or toroidal).
In this paper we will consider only the even sector.\\
We write the (even) metric perturbation as \be h_{\mu \nu}=\left(
\begin{tabular}{cc} $h_{AB} Y$ & $h_A Y,_a$ \\  Symm
& $r^2(K Y \gamma_{ab}+G Y,_{a:b})$ \end{tabular} \right),
\label{evenmetp}\ee and the (even) matter perturbation as \be
\Delta t_{\mu \nu}=\left(
\begin{tabular}{cc} $\Delta t_{AB} Y$ & $\Delta t_A Y,_a$ \\  Symm
& $r^2 \Delta t^1 Y \gamma_{ab}+\Delta t^2 Z_{ab}$
\end{tabular} \right). \ee
Thus the metric perturbation is defined by a symmetric two-tensor,
a two-vector and two scalars, \be h_{AB},\qquad h_A,\qquad
G,\qquad K, \label{baremet} \ee and the matter perturbation is
defined by the same, \be \Delta t_{AB}, \qquad \Delta t_A, \qquad
\Delta t^1, \qquad \Delta t^2. \label{baremat}\ee When we write
down Einstein's equations for this metric and matter tensor, we
identify the left and right hand side coefficients of the scalar,
vector and tensor bases given in
(\ref{vecbases}),(\ref{tenbases}), and these are our evolution
equations for the perturbation. The great simplification is that
these equations are in terms of two-dimensional objects $h_{AB}$
etc, and their derivatives, defined on the manifold $M^2$. This
makes calculating the perturbation equations much easier as the
connections used in calculating derivatives e.g. $h_{A|B}$ are
those defined on the background.\\
It is important to note that for $l=0$ all of the basis objects
given in (\ref{vecbases}),(\ref{tenbases}) vanish except for
$Y\gamma_{ab}$, and for $l=1$ $Z_{ab}$ and $S_{(a:b)}$ vanish.
Thus some equations do not exist in these cases, and we must
consider $l=0$ and $l=1$ separately from the more general $l\geq
2$ case.

\subsection{Gauge invariance.}

Two space-times are identical if they only differ by a
diffeomorphism \cite{wald} (we take the passive view of a
diffeomorphism as a coordinate transformation). There is a danger
that if you add a ``perturbation'' \be
g_{\mu\nu}=\w{g}_{\mu\nu}+h_{\mu\nu}, \ee you are in fact still
looking at the same space-time after undergoing a coordinate
transformation, rather than after being perturbed in a physically
meaningful way. To escape this problem we must only interest
ourselves in those objects which do not change under an
infinitesimal coordinate (gauge) transformation. These are called
gauge invariants, and are the true measure of a physically
meaningful perturbation. (More precisely these are identification
gauge
invariant, see Stewart and Walker \cite{StewartWalker} for details).\\
The vector field generating a gauge transformation has an even/odd
parity decomposition as  \be \bmt{\xi}=\xi_A Y dx^A+\xi Y,_a dx^a.
\ee The gauge change induced on the tensor $h_{\mu\nu}$ is \be
h_{\mu\nu}\rightarrow
\ov{h}_{\mu\nu}=h_{\mu\nu}+\mathcal{L}_{\vec{\xi}}\w{g}_{\mu \nu},
\label{defgt}\ee where an over-bar represents gauge transformed
objects, and where \be \mathcal{L}_{\vec{\xi}}\w{g}_{\mu
\nu}=\nabla_\mu \xi_\nu + \nabla_\nu \xi_\mu, \ee with
$\nabla_\mu$ the covariant derivative associated with the
\emph{background} metric. Thus we can write down how all of the
perturbation objects in $h_{\mu\nu}$ and $\Delta t_{\mu\nu}$
transform \cite{GS}.

Now we construct the gauge invariants, that is we take linear
combinations of (\ref{baremet}), (\ref{baremat}) to form objects
which do not change under a gauge transformation. These are
\begin{subequations}\label{gis}\be\left.\begin{array}{rcl} k_{AB}&=&h_{AB}-(p_{A|B}+p_{B|A})
\\ k&=&K-2\w{v}^A p_A \end{array}\right\}(metric) \\
\left.\begin{array}{rcl} T_{AB}&=&\Delta
t_{AB}-\w{t}_{AB|C}p^C-\w{t}_A^{\phantom{A}C}p_{C|B}-\w{t}_B^{\phantom{B}C}p_{C|A}
\\ T_A&=&\Delta t_A-\w{t}_A^{\phantom{A}C}p_{C}-r^2(\w{t}^a_{\phantom{a}a}/4)G,_A \\ T^1&=& \Delta t^1
-(p^C/r^2)(r^2
\w{t}^a_{\phantom{a}a}/2),_C+l(l+1)(\w{t}^a_{\phantom{a}a}/4)G \\
T^2&=& \Delta t^2-(r^2 \w{t}^a_{\phantom{a}a}/2)G
\end{array}\right\}(matter) \ee\end{subequations}
where $p_A=h_A-\f{1}{2}r^2 G,_A$ and $\w{v}^A=\f{r^{|A}}{r}$. The
perturbation evolution equations are then recast entirely in
terms of the gauge invariants. We give these equations in Appendix A.\\
There is an especially useful gauge choice we can make, called the
Regge-Wheeler or longitudinal gauge. This consists of transforming
to a specific coordinate system via \be
\bmt{\xi}=\left(h_A-\f{r^2}{2}G,_A\right) Y dx^A+\f{r^2}{2} G Y,_a
dx^a, \label{RWg} \ee in which $h_A=G=0=p_A$. Since we are
measuring gauge invariants we are free to make this
transformation, the benefit being that in this gauge the bare
perturbations (\ref{baremet}),
(\ref{baremat}) and the gauge invariants match.\\
When $l=0,1$, we cannot construct a set of gauge invariant objects
like those given above for the same reason that there are less
equations in these sectors (see Appendix A): the vanishing of some
or all of the bases given in (\ref{vecbases}),(\ref{tenbases}).
Thus for $l=1$ modes we can at best construct only partially gauge
invariant objects, and for $l=0$ all remnants of gauge invariance
are lost.
\\
\\
Finally we must consider what to measure on the relevant surfaces
in order to test for stability. Following Chandrasekhar
\cite{chand}, we use the Weyl scalars to measure the flux of
energy of the perturbations. For a detailed discussion on how the
Weyl scalars relate to the Zerilli, Regge-Wheeler, Moncrief etc.
scalars, see Lousto \cite{Lousto}. Stewart and Walker
\cite{StewartWalker} showed that the only Weyl scalars which are
both identification gauge invariant, which is the sense described
above, and tetrad gauge invariant (independent of the choice of
null tetrad with which the Weyl scalars are defined), are the
Petrov type-N terms. Furthermore, they are only tetrad and
identification gauge invariant if the background is type-D or
conformally flat. Since Vaidya space-time is spherically
symmetric, and therefore type-D, this means there are two fully
(tetrad and identification) gauge invariant Weyl scalars (for
$l\geq 2$), $\delta \Psi_0$ and $\delta \Psi_4$. These scalars
represent pure transverse gravitational waves propagating in the
radial inward (respectively outward) null directions of a
spherically symmetric background.\\
With a particular choice of null tetrad these scalars can be given
as \be \delta \Psi_0=\f{1}{2r^2}\w{\ell}^A \w{\ell}^B k_{AB}
(\w{w}^a \w{w}^b Y_{:ab}), \quad \delta \Psi_4=\f{1}{2r^2} \w{n}^A
\w{n}^B k_{AB} (\w{w}^{*a} \w{w}^{*b} Y_{:ab}), \ee where
$\w{\ell}^\mu,\w{n}^\mu,\w{m}^\mu=r^{-1} \,
\w{w}^\mu(\theta,\phi)$ and $\w{m}^{*\mu}$ are a null tetrad of
the background and the $*$ represents complex conjugation (see,
for example, Nolan \cite{brien2}). Tetrad gauge invariance of
these objects must be carefully interpreted. In the type-D
background, there is an obvious choice of null tetrad: we take
$\w{\ell}^\mu,\w{n}^\mu$ to be the the principal null directions
of the Weyl tensor, and take $\w{m}^\mu$ and its conjugate to be
unit space-like vectors orthogonal to $\w{\ell}^\mu,\w{n}^\mu$ to
complete the tetrad. Then $\delta\Psi_0$ and $\delta\Psi_4$ are
identification gauge invariant and also tetrad gauge invariant
with respect to any {\em infinitesmal} Lorentz transformation of
the tetrad, and also with respect to any finite null rotation that
leaves the directions of $\w{\ell}^\mu,\w{n}^\mu$ fixed. However
these scalars {\em are not} preserved under the {\em finite}
boost-rotations \[ \w{\ell}^\mu\to A\w{\ell}^\mu,\quad
\w{n}^\mu\to A^{-1}\w{n}^\mu,\quad \w{m}\to e^{i\omega}\w{m},\quad
A>0,\omega\in\mathbb{R}\] under which they rescale as
\[\delta\Psi_0\to A^2\delta\Psi_0,\quad\delta\Psi_4\to
A^{-2}\delta\Psi_4.\]For the sake of boundary conditions however
we must take this scale covariance into account, (see Beetle and
Burko \cite{BeetleBurko} for a generalized discussion), and thus
to first order our ``master'' function will be \cite{brien2} \be
\delta P_{-1}=|\delta \Psi_0 \delta \Psi_4|^{1/2}. \ee For $l=0,1$
the angular part of $\delta \Psi_0, \delta \Psi_4$ is zero, and
thus these scalars vanish in these sectors.

\section{Minkowski region.}

For completeness our analysis must include the region of flat
space proceeding the Vaidya region. In our global, self-similar
coordinates $(x,r)$, Minkowski space-time is given by \be
g_{\mu\nu}dx^\mu dx^\nu=-r^2 dx^2+2r(1-x) dx dr+x(2-x) dr^2+r^2
d\Omega^2,\qquad t_{\mu\nu}dx^\mu dx^\nu =0.\ee We seek to
identify only that class of perturbations which are finite on the
portion of the axis $r=0, v<0$, and on $\mathcal{N}$, the surface
marking the transition between flat space-time and Vaidya
space-time, which we call the ``threshold''. $\mathcal{N}$ is the
past null cone of the origin $(x,r)=(0,0)$. We consider the case
where the perturbed space-time is still empty, that is the matter
perturbation is zero. Thus the right hand sides of all
perturbation equations is zero.

\subsection{$l\geq 2$ modes, Minkowski.}

The perturbation equations for Minkowski background are
\begin{eqnarray}
2\w{v}^C(k_{AB|C}-k_{CA|B}-k_{CB|A}-\w{g}_{AB}k_{CD}^{\phantom{CD}|D}+\w{g}_{AB}k_{CD}\w{v}^D)-\f{l(l+1)}{r^2}k_{AB}\nonumber
\\
-\w{g}_{AB}\left(2k,_C^{\phantom{,C}|C}-\f{(l-1)(l+2)}{r^2}k\right)+2(\w{v}_Ak,_B+\w{v}_Bk,_A+k,_{A|B})=0,\nonumber\\
(\w{v}^C\w{v}^D+\w{v}^{C|D})k_{CD}=0, \qquad
k,_A-k_{AC}^{\phantom{AC}|C}=0, \qquad k_A^{\phantom{A}A}=0,
\end{eqnarray} where $\w{v}^A=r^{|A}/r$.\\
We write these equations in $(x,r)$ coordinates and perform a
Mellin transform (\ref{mell}) over $r$. There are four unknowns,
which, after the Mellin transform, we denote
\be k_{AB}=\left(%
\begin{array}{cc}
  r^{s+1}A_s(x) & r^s B_s(x) \\
  r^s B_s(x) & r^{s-1} C_s(x) \\
\end{array}%
\right), \qquad k=r^{s-1} K_s(x). \ee We define a new variable
$D=B-xA$, where for brevity we have dropped the subscript $s$. The
scalars we wish to measure then become \be
\delta\Psi_0=-2r^{s-3}D,\qquad \delta\Psi_4=\f{1}{2}r^{s-3}(2A+D),
\qquad \delta P_{-1}=|\delta \Psi_0 \delta \Psi_4|^{1/2}.
\label{scmink}\ee We can decouple the system to get an ode in $D$,
\be x(x-2)D''+2(1+s+x-s x)D'-(l+l^2+s-s^2)D=0, \ee which is
hypergeometric, \be
z(1-z)D''(z)+(\gamma-(\alpha+\beta+1)z)D'(z)-\alpha \beta D(z)=0,
\ee with $\alpha=1+l-s$, $\beta=-l-s$, $\gamma=-1-s$, and $z=x/2$.
There are two solutions near the axis $x=-\infty$ \cite{SpecFunc},
denoted by the subscript {\sc a},  \be
&&\phantom{}_{\textrm{{\sc a}}}D_1=(-z)^{s-l-1} \phantom{}_2F_1(1+l-s,3+l;2+2l;z^{-1}), \\
&&\phantom{}_{\textrm{{\sc a}}}D_2=-\ln(-z)
\phantom{}_{\textrm{{\sc a}}}D_1+(-z)^{s+l}(1+O(z^{-1})), \ee
since $\alpha-\beta=2l+1 \in \mathbb{Z}$.

If we form the general solution \be D=d_1
\,\phantom{}_{\textrm{{\sc a}}}D_1+d_2 \, \phantom{}_{\textrm{{\sc
a}}}D_2, \ee and from this solutions for $A,B,C$ and $K$, we find
to leading order near $x=-\infty$ ($r=0$), \be \delta \Psi_{0,4}
\sim d_1 r^{l-2} (1+O(r))+d_2 r^{-l-3} (1+O(r)). \ee Thus in order
to make $\delta \Psi_{0,4}$, and hence $\delta P_{-1}$, regular at
the axis, we need to set $d_2=0$ in the general solution for $D$.

Now we allow only the first solution for $D$ to evolve up to the
past null cone of the origin $\mathcal{N}$, the threshold. When we
do so we use the nature of the hypergeometric equation to write
the acceptable solution at the regular axis as a linear
combination of the two solutions on $\mathcal{N}$. That is to say,
near $x=0$ this solution has the form \be \phantom{}_{\textrm{{\sc
a}}}D_1=d_3\, \phantom{}_{\textrm{{\sc t}}}D_1+d_4\,
\phantom{}_{\textrm{{\sc t}}}D_2, \label{sumD} \ee where
$\phantom{}_{\textrm{{\sc t}}}D_1,\phantom{}_{\textrm{{\sc
t}}}D_2$ are two naturally arising linearly independent solutions
of the hypergeometric equation near $x=0$. In finding these
solutions, the relation between $\alpha,\beta$ and $\gamma$ is key
so we must
consider two cases: $s\in \mathbb{Z}$ and $s \notin \mathbb{Z}$.\\
\\
The more straightforward case is when $s\notin \mathbb{Z}$. Then
$1-\gamma \notin \mathbb{Z}$ and we can take  \be
\phantom{}_{\textrm{{\sc t}}}D_1=F(\alpha,\beta;\gamma;z), \qquad
\phantom{}_{\textrm{{\sc t}}}D_2=z^{1-\gamma}
F(\alpha-\gamma+1,\beta-\gamma+1;2-\gamma;z), \label{nninZhyp}\ee
and (\ref{sumD}) holds with \cite{Bate} \be
d_3=\f{\Gamma(1-\gamma) \Gamma(\alpha-\beta+1)}{\Gamma(1-\beta)
\Gamma(\alpha-\gamma+1)}, \qquad
d_4=-\f{\Gamma(\gamma)\Gamma(1-\gamma)\Gamma(\alpha-\beta+1)}{\Gamma(2-\gamma)
\Gamma(\gamma-\beta)\Gamma(\alpha)}e^{i \pi (\gamma-1)}.
\label{dconstants}\ee Again the solutions for $A,B,C$ and $K$ can
be recovered from these expressions for $D$.

When we calculate the scalars due to $\phantom{}_{\textrm{{\sc
t}}}D_1,\phantom{}_{\textrm{{\sc t}}}D_2$ near $x=0$ (and away
from the singularity ar $r=0$) we find \be \delta \Psi_0 \sim d_3
(1+O(x))+d_4 x^{s+2} (1+O(x)), \quad \delta \Psi_4 \sim d_3
(1+O(x))+d_4 x^{s-2} (1+O(x)). \ee These two scalars and $\delta
P_{-1}$ will be finite on $x=0$ iff \be Re(s)>2, \quad s \notin
\mathbb{Z}.\ee We can also see this with a general argument: the
other singular point of the hypergeometric equation is at $z=1$,
or $x=2$. This is the future null cone of the origin in Minkowski
space-time. We would expect solutions to be regular here too, and
from the transformation \be
\phantom{}_2F_1(\alpha,\beta;\gamma;z)=(1-z)^{\gamma-\alpha-\beta}
\phantom{}_2F_1(\gamma-\alpha,\gamma-\beta;\gamma;z) \ee we
require $Re(\gamma-\alpha-\beta)=Re(s-2)>0$.\\
\\
The case $s \in \mathbb{Z}$ is cumbersome and we will only
summarize the results here. This case is again split according to
the sign of $1-\gamma \in \mathbb{Z}$, however there are no
solutions for $1-\gamma \leq 0$ which are finite on $\mathcal{N}$
and thus we only present here the solutions for $1-\gamma>0$ ($s+2>0$).\\
If $\gamma=1-m$ where $m$ is a natural number, and $\alpha,\beta$
are different from the numbers $0,-1,-2,\ldots,1-m$, then there
are two linearly independent solutions near $x=0$ given by
\cite{GradRyzh} \be \phantom{}_{\textrm{{\sc
t}}}D_1&=&z^{1-\gamma}
F(\alpha-\gamma+1,\beta-\gamma+1;2-\gamma;z), \nonumber \\
\phantom{}_{\textrm{{\sc t}}}D_2&=&\ln (z) \,
\phantom{}_{\textrm{{\sc t}}}D_1- \sum_{n=1}^m
\f{(n-1)!(-m)_n}{(\gamma-\alpha)_n(\gamma-\beta)_n}
z^{m-n}+\sum_{n=0}^{\infty} \f{(\alpha+m)_n(\beta+m)_n}{(1+m)_n
n!} [h^*(n)-h^*(0)] z^{m-n}, \label{ninZ} \ee where \be
h^*(n)=\psi(\alpha+m+n)+\psi(\beta+m+n)-\psi(1+m+n)-\psi(1+n), \ee
and $\psi=\psi^{(0)}=\frac{\Gamma'}{\Gamma}$ is the Digamma
function. The scalars $\delta \Psi_{0,4}$ and $P_{-1}$ will be
finite on $x=0$
due to these solutions if $Re(s)\geq 2$.\\
If, however, $\alpha$ or $\beta$ is equal to one of the numbers
$0,-1,-2,\ldots,1-m,$ then the solution given above loses meaning;
this will occur for $s=l+m$, where $m$ is a natural number ($\geq
1$). In this case two linearly independent solutions are \be
&&\phantom{}_{\textrm{{\sc t}}}D_1=F(\alpha,\beta;\gamma;z), \\
&&\phantom{}_{\textrm{{\sc
t}}}D_2=F(\alpha-\gamma+1,\beta-\gamma+1;2-\gamma;z).\ee The
scalars due to these solutions \emph{will} diverge on $x=0$.
Therefore we must \emph{not} consider the modes $s=l+m$ when
summing to find the general solution. This does not challenge the
generality of the result however, as all multipole mode numbers
$l$ ($\geq 2$) will still be counted.\\
\\
Thus we have found the largest class of perturbations that will
have finite flux on the axis and on the threshold: all $s \in
\mathbb{C}$ such that $Re(s)\geq 2$, except for $s=l+m$ where $m$
is a natural number. These solutions will be matched across $x=0$
into the Vaidya space-time, and then allowed evolve up to the
Cauchy horizon.

\subsection{$l=1$ mode, Minkowski.}

While the result in this sector is well known, for the sake of
completeness we briefly give the analysis in terms of the
formalism defined above.

In this sector not all the perturbation equations apply, and we
can only define partially gauge invariant objects. Looking at
(\ref{evenmetp}), for $l=1$ we find $Y \gamma_{ab}=-Y,_{a:b}$ and
thus \be h_{\mu\nu}dx^{\mu}dx^{\nu}=\ldots+ r^2(K-G) Y \gamma_{ab}
dx^a dx^b. \ee If we let $H=K-G$, then the gauge transformation of
$H$ generated by $\xi_\mu dx^\mu=\xi_A Y dx^A+ \xi Y,_a dx^a$ is
\be \ov{H}-H=\ov{K}-K-(\ov{G}-G)=2 \xi/r^2-2\w{v}^A \xi_A. \ee Now
we rename $H$ as $K$, and we have effectively set $G=0$ with $K$
now transforming as $H$ given above. Thus $p_A=h_A$, and we are
left with this sensitivity to the angular part of the gauge
transformation, \be k_{AB}& \rightarrow &
k_{AB}+\left[r^2(\xi/r^2),_A\right]_{|B}+\left[r^2(\xi/r^2),_B\right]_{|A}\\
k& \rightarrow & k+2\xi/r^2+2\w{v}^A r^2 (\xi/r^2),_A.
\label{ktrans} \ee We can, however, use this to our advantage.
Following Sarbach and Tiglio \cite{Sarbach}, we look to transform
into a coordinate system in which $k_A^{\ph{A}A}=0$. To do this we
choose $\xi$ such that \be
\left[r^2(\xi/r^2),_A\right]^{|A}=-k_A^{\ph{A}A}.\ee Then we are
free to make further gauge transformations, provided \be
\left[r^2(\xi/r^2),_A\right]^{|A}=0. \label{gcl1}\ee Thus we can
reinstate the second scalar perturbation equation, $k_A^{\ph{A}A}=0$, as a gauge choice.\\
Using this, we split the tensor perturbation equation into its
trace and trace-free parts. Then the set of equations for $l=1$ is
given by \be
2\w{v}^C(k_{AB|C}-k_{CA|B}-k_{CB|A})-\f{2}{r^2}k_{AB}+2(\w{v}^A
k,_B+\w{v}^B k,_A+k,_{A|B})-\w{g}_{AB}k,_{C}^{\ph{C}|C}=0, \nonumber \\
2\w{v}^C \w{v}^D k_{CD}=k,_{C}^{\ph{C}|C}+2\w{v}^C k,_C \qquad
k_{AC}^{\ph{AC}|C}=k,_A \hspace{1in} \ee where the first equation
is the trace-free part of the tensor equation, and the second is
its trace.

Now we must consider what to solve the equations for. The scalars
$\delta\Psi_{0,4}$ given before have an angular dependence which
is zero for $l=1$, and thus they vanish in this sector. The other
options for true scalars to measure are $k_A^{\ph{A}A}$ and $k$,
however we have chosen a gauge in which the trace of $k_{AB}$ is
zero, and thus the only scalar left to measure is $k$. We will
show this scalar is pure gauge, that is there is enough residual
gauge freedom to transform into a gauge in which $k=0$. In this
gauge, all the components of $k_{AB}$ are also zero, and thus this
perturbation sector is empty.

For convenience we look at the perturbation equations in
orthogonal coordinates, \be ds^2=-dt^2+dr^2+r^2 d\Omega^2, \ee and
in this coordinate system we label components \be k_{AB}=\left(%
\begin{array}{cc}
  A(t,r) & B(t,r) \\
  B(t,r) & A(t,r) \\
\end{array}%
\right), \label{compkh} \ee since $k_{AB}$ is symmetric and
trace-free. Now when we look at the perturbation equations, the
trace equation gives $A$, and one of the trace-free equations
gives $B$, in terms of $k$ and its derivatives, \be
A(t,r)&=&\frac{r}{2} \Big( 2 k,_r+ r (k,_{rr}-k,_{tt})\Big),
\nonumber \\ B(t,r)&=&\frac{r}{2} \Big( 2 k,_t+r^2
(k,_{tt}-k,_{rr}),_t \Big). \ee The remaining equations give \be r
\left(\Box k\right),_r= -4 \left(\Box k\right), \ee which we solve
as \be \Box k =f(t)\, r^{-4}. \ee $k$ should satisfy this equation
with initial data $k(0,r)=\alpha(r)$, $\dot{k}(0,r)=\beta(r)$
satisfying $\alpha, \, \beta \in C^1$ and $k(t<0,r=0)$ finite.
This implies $f(t)=0$, and thus $k$ solves the homogeneous wave
equation \be \Box k=0. \ee $k$ however is not gauge invariant.
Looking at (\ref{ktrans}), we see $k \rightarrow \ov{k}=k+\eta$
where $\eta=2\xi/r^2+2\w{v}^A r^2 (\xi/r^2),_A.$ Thus \be \Box k&
\rightarrow & \Box \ov{k}=\Box k+\Box \eta,\ee but $\Box \eta=0$
(provided (\ref{gcl1}) holds), and thus $k$ and $\eta$ satisfy the
same equation. Therefore we can choose $\eta=-k$, and thus we can
always transform to a gauge in which $k=0$. In this gauge we also
have $k_{AB}=0$, and thus the entire $l=1$ perturbation is pure
gauge.

\subsection{$l=0$ mode, Minkowski.}

The $l=0$ mode represents a spherically symmetric perturbation. As
we are considering zero matter perturbation, we have the perturbed
space-time is spherically symmetric vacuum, and thus Birkhoff's
Theorem applies, that is the perturbed space-time is Schwarschild.
We will recover the specifics using our perturbation formalism
described above.

For $l=0$, $Y,_a=0$ and thus $h_A=G=0$. We cannot form gauge
invariants and thus use $K$ and $h_{AB}$, which are fully
dependent on gauge transformations $\xi_\mu dx^\mu=\xi_A dx^A$ as
\be h_{AB} &\rightarrow& h_{AB}-(\xi_{A|B}+\xi_{B|A}),\nonumber
\\ K &\rightarrow& K-2\w{v}^A \xi_A.
\ee We can transform to a gauge in which $K=h^A_{\ph{A}A}=0$ by
choosing $\xi_A$ such that \be
\xi_A^{\ph{A}|A}=\tfrac{1}{2}h^A_{\ph{A}A}, \qquad \w{v}^A
\xi_A=\tfrac{1}{2} K. \ee In $(t,r)$ coordinates, this means
making a gauge transformation $\xi_r=\tfrac{1}{2} r K$ and
$\xi_{t,t}=\xi_{r,r}-\tfrac{1}{2}h_A^{\ph{A}A}$. Further
transformations preserving $K=h^A_{\ph{A}A}=0$ must be of the form
$\xi_A=f(r) \delta_A^{\,t}$ for any arbitrary function $f(r)$.
Thus of the remaining perturbation terms, $A(t,r)$ is gauge
invariant, whereas $B(t,r) \rightarrow B(t,r)-f'(r)$, where \be h_{AB}=\left(%
\begin{array}{cc}
  A(t,r) & B(t,r) \\
  B(t,r) & A(t,r) \\
\end{array}%
\right).\ee The perturbation equations reduce to $A,_t=B,_t=0$ and
$r A,_r=-A$. Since $B$ is a function of $r$ alone, we can choose
$f(r)$ to set $B=0$. Thus the only perturbation term which cannot
be gauged away is \be A=\f{c}{r}. \ee Renaming the constant
$c=2m$, and noting $ \left(1-2m/r\right)^{-1}\approx
\left(1+2m/r\right) $ for $m$ small (which is the case in this
linear model), we have recovered the Schwarzschild line element.
There is an intrinsic singularity on the axis $r=0$, and thus this
solution contradicts our initial data; therefore there is no $l=0$
perturbation.

\section{Vaidya region.}

In this section we describe metric and matter perturbations of the
Vaidya solution given in (\ref{Vmetric}),(\ref{Vmatter}). Initial
conditions for this problem are that any perturbations coming from
the region of flat space preceding the threshold should be finite
on, and match continuously across, this surface.\\
As before, we will need to split the analysis into the $l=0,l=1,$
and $l \geq 2$ cases.

\subsection{$l \geq 2$ modes, Vaidya.}

We consider the case where the perturbed space-time is that of a
null dust, and thus has matter tensor \be t_{\mu
\nu}=(\w{\rho}+\delta \rho)(\w{\ell}_\mu+\delta
\ell_\mu)(\w{\ell}_\nu+\delta \ell_\nu), \label{pertnulldust} \ee
where $\w{\rho}+\delta \rho$ is the energy density and
$\w{\ell}_\mu+\delta \ell_\mu$ is the null vector of the perturbed
space-time. As before we decompose these perturbation objects in
terms of the spherical harmonics. Since we are measuring the gauge
invariants given in (\ref{gis}), we can examine in any gauge we
choose. For convenience, we will use the Regge-Wheeler (RW) gauge
given in (\ref{RWg}). Thus we can write the matter gauge
invariants as
\be T_{AB}=\left(%
\begin{array}{cc}
  r^2 \delta\rho-\f{\lambda\partial_x\Gamma}{4\pi r} & r x \delta\rho-\f{\lambda\partial_r\Gamma}{8\pi r}-\f{\lambda x \partial_x\Gamma}{8\pi r^2} \\
  \textrm{Symm} & x^2 \delta\rho-\f{\lambda x \partial_r\Gamma}{4\pi r^2} \\
\end{array}%
\right), \qquad
T_A=\left(%
\begin{array}{c}
  \f{-\lambda \Gamma}{8\pi r} \\
  \f{-\lambda x \Gamma}{8\pi r^2} \\
\end{array}%
\right), \qquad T^1=T^2=0.  \label{mattpert}\ee We define $\Gamma$
as the perturbation in the null coordinate; that is we can define
the null coordinate of the perturbed space-time as
$V=v-\Gamma(x,r) Y(\theta,\phi)$. It follows from the conservation
equations that the null vector of the perturbed space-time is
$\w{\ell}_\mu+\delta
\ell_\mu=-\p_\mu V$.\\
We write out the perturbation equations as given in appendix A,
and as before perform a Mellin transform over $r$. There are now
six unknowns,
\be k_{AB}=\left(%
\begin{array}{cc}
  r^{s+1}A_s(x) & r^s B_s(x) \\
  r^s B_s(x) & r^{s-1} C_s(x) \\
\end{array}%
\right), \qquad k=r^{s-1} K_s(x), \qquad \Gamma=r^s G_s(x), \qquad
\delta \rho=r^{s-3} M_s(x). \label{mellofpert}\ee A constraint
equation defining $\delta \rho$ decouples and so we are left with
five unknowns in the evolution system. We use the scalar equation
(\ref{tracek}) to remove one of the metric variables, and so we
can reduce the set of equations to a four dimensional first order
linear system. We give this system in Appendix B.
\\
As before (dropping the subscript $s$), we define the variable
$D=B-x A$, and the scalars to measure become \be \delta
\Psi_0=\f{2r^{s-3}D}{\lambda x-1}, \quad \delta
\Psi_4=\f{r^{s-3}(2A+(1- \lambda x)D)}{2}, \quad \delta
P_{-1}=|\delta \Psi_0 \delta \Psi_4|^{1/2}. \ee From our system of
equations we can decouple a second order ordinary differential
equation for $D$, \be D''(x)+p(x) D'(x)+q(x) D(x)=0,
\label{deqn}\ee where \be p(x)&=&-\f{s+1}{x}+\f{2
\lambda}{1-\lambda
x}+\f{s-3-(s-6)\lambda x}{\lambda x^2-x+2}, \\
q(x)&=&\f{(\lambda x-1)^2(l+l^2+s^2(\lambda
x-1))+\lambda(x-2+6\lambda x-\lambda x^2)-s(\lambda
x-1)(1+\lambda(-2+x(2\lambda x-3)))}{x (\lambda x-1)^2(\lambda
x^2-x+2)}. \ee The regular singular point at $x=0$ is the past
null cone of the origin of coordinates, and is the surface over
which we move from flat space-time to Vaidya space-time (the
threshold). When $\lambda < 1/8$ the space-time has the structure
given in Figure 1, and there are two surfaces which are regular
singular points of the above equation: the first is the Cauchy
horizon at $x=x_c$ and the second is the second future similarity
horizon (SFSH) $x=x_e$, the first and second zeroes of $\lambda
x^2-x+2$ respectively. Finally $x_a=1/\lambda$ is a regular
singular point describing the apparent horizon.

We will begin the analysis by examining the threshold $x=0$ and
ensuring our initial conditions are met. As the analysis will
differ depending on whether $s$ is an integer or not, we will
consider these two cases separately. Then we will allow the
acceptable solutions to evolve up to the Cauchy horizon, and then
on to the SFSH.

\subsubsection{Threshold, $s \not\in \mathbb{Z}$ .}

We can use the method of Frobenius to describe solutions near
regular singular points, and we present these using the $P$-symbol
notation \cite{SpecFunc}. For the first four columns in the
$P$-symbol, the first row's entry denotes the location of the
regular singular point, the second and third rows' denote the
leading order exponents of two infinite power series solutions at
that point. If these exponents do not differ by an integer, then
these series are two linearly independent solutions for $D$ near
that point; if the exponents do differ by an integer, then a
logarithmic term may be introduced. Thus, for $\lambda < 1/8$, \be
P\left\{\begin{array}{ccccc}
  0 & x_c & x_e & x_a & \\
  0 & 0 & 0 & 1 &  ; x \\
  s+2 & \f{1}{2}\left(s-4+\f{s}{\sqrt{1-8\lambda}}\right) & \f{1}{2}\left(s-4-\f{s}{\sqrt{1-8\lambda}}\right) & 2 & \\
\end{array}\right\} \label{psymbol}.
\ee As this is a second order equation coming from a fourth order
system, there must be two solutions with $D=0$. We find these by
setting $D=0$ in the system, which simplifies the equations
greatly. Thus we can find the set of exact solutions corresponding
to $D=0$, which we call solutions $III$ and $IV$ and are valid
everywhere and irrespective of whether $s$ is an integer or not:
\be
\begin{array}{cc|c} &
III & IV \\
A= & 2g_0 \lambda x^s & (s-1)k_0 x^{s-2}-\f{(l^2+l-2)}{2}k_0 x^{s-1} \\
D= & 0 & 0 \\
K= & 0 & k_0 x^{s-1} \\
G= & g_0 x^s & 0, \label{deqzer}
\end{array}
\ee where $k_0,g_0$ are arbitrary constants. The full set of
solutions are found using system methods, as this approach has the
benefit of solving for all four solutions at once, with the
required accuracy found by simply taking further terms in the
series. These system methods are
described in Appendix C, and we briefly outline their application here:\\
The system of perturbation equations can be written in the form
\be Y'=\f{1}{x^2}\left(J+\sum_{m=1}^{\infty}A_m x^m\right)Y,\ee
where $Y=(A,D,K,G)^T$, and $J \neq 0$. Thus $x=0$ is an irregular
singular point. We find that $J$ has eigenvalue $0$ multiplicity
four. $J$ cannot be diagonalised and we use Theorem
\ref{theorem:thfour} given in Appendix C to remove off-diagonal
terms, which effectively reduces the order of the singularity to a
regular singular point. Now the leading order coefficient matrix
has eigenvalues $0,s,s,s-2$, and so we apply Theorem
\ref{theorem:thtwo} twice to reduce the eigenvalues to $0$ and
$s-2$ multiplicity three. Finally we apply Theorem
\ref{theorem:thone} to obtain the following.\\
To leading order, we find a full set of linearly independent
solutions with asymptotic behaviour (as $x\rightarrow 0$)

\be \begin{array}{cc|c|c|c} &
I & II & III & IV \\
A= & O(1) & O(x^{s-2})& O(x^s) & O(x^{s-2}) \\
D= & O(1) & O(x^{s+2})& 0 & 0 \\
K= & O(1) & O(x^{s+1})& 0 & O(x^{s-1}) \\
G= & O(1) & O(x^{s+2}) & O(x^s) & 0
\end{array} \label{leadordthr}
\ee Solutions $I$ and $II$ correspond with the Minkowski solutions
(matching across $x=0$ is dealt with in the next subsection), and
$III$ and $IV$ are the $D=0$ solutions given in (\ref{deqzer}) .

Taking $A$ and $D$ to be a linear combination of these solutions,
we calculate the leading order of the scalars near $x=0$ as \be
\delta \Psi_0 \sim O(1)+O(x^{s+2}), \quad \delta \Psi_4 \sim
O(1)+O(x^{s-2})+O(x^s), \ee (for brevity we have left out the
constants of combination). Thus for our master function $\delta
P_{-1}$ to be finite on the threshold due to the Vaidya solutions,
we find we must maintain the same constraint as that coming from
the flat space solutions, that is \be Re(s)>2, \quad s \notin
\mathbb{Z}. \ee

\subsubsection{Threshold, $s \in \mathbb{Z}$.}

When $s$ is an integer the system methods break down for the
following reason: the eigenvalues of the leading order coefficient
matrix of the regular singular point at $x=0$ are $0$ and $s-2$
multiplicity three. These differ by an integer, and thus they must
be repeatedly reduced until they are equal. However, each time we
reduce an eigenvalue we must diagonalise the leading order
coefficient matrix, which prevents us from simply reducing the
eigenvalue (the unspecified number) $s-2$ times.\\
Instead, we use the ordinary differential equation for $D$ to
write the solutions near $x=0$ as, using the fact that $s+2>0$
(since the flat space solutions constrained $s \geq 2$), \be D=d_5
\sum_{m=0}^{\infty} A_m x^{m+s+2}+d_6 \left\{ k \ln x
\sum_{m=0}^{\infty} A_m x^{m+s+2}+\sum_{m=0}^{\infty} B_m x^m
\right\}, \label{dsolsint}\ee with $d_5,d_6$ constants and $k$
possibly zero.

We use this solution as an inhomogeneous term to solve for the
other variables by integration, and we find \be G&\sim&g_0 x^s-d_1
\sum_{m=0}^{\infty} \f{A_m x^{m+s+2}}{m+2}-d_2
\left\{\sum_{m=0}^{\infty} \f{B_m x^m}{m-s}+k \sum_{m=0}^{\infty}
\f{A_m x^{m+s+2}}{m+2}\left(\ln x -\f{1}{m+2}
\right) \right\}, \\
K&\sim&k_0 x^{s-1}+d_1 \left\{\sum_{m=0}^{\infty} \f{A_m
x^{m+s+1}}{m+2} \left(-2(m+s+2)+x(1-s-2\lambda)\right)
\right\}+d_2 \left\{\sum_{m=0}^{\infty}B_m x^m \left(
\f{1-s-2\lambda}{m-s+1}-\f{2 m}{(m-s)x}\right)\right. \nonumber \\
&&\left. +k \sum_{m=0}^{\infty} A_m x^{m+s+1}
\left(\f{-2}{m+2}-\f{2(m+s+2)}{m+2}\left(\ln
x-\f{1}{m+2}\right)+\f{x(1-s-2\lambda)}{m+3}\left(\ln
x-\f{1}{m+3}\right)\right)\right\}, \\
A&\sim&(s+l^2+l-1) x^{-1} D+2\lambda G+(s^2-1)x^{-1}K-s K'.
 \ee
Since both $D$ and $K$ have $O(1)$ terms, we see there is an $A$
solution which diverges on the threshold like $x^{-1}$. This
divergent term cannot be switched off, for the following reason:\\
\\
On the axis there were two solutions for $D$, which we denoted \be
\phantom{}_{\textrm{{\sc a}}}D_1 \sim x^{s-l-1}, \quad
\phantom{}_{\textrm{{\sc a}}}D_2 \sim x^{s+l}. \ee The scalars
$\delta \Psi_{0,4}$ due to the first solution went like $r^{l-2}$,
whereas the second solution gave $\delta \Psi_{0,4} \sim
r^{-l-3}$. Thus we needed to switch off the divergent term in the
general solution, $D|_{x=-\infty}=d_1 \,\phantom{}_{\textrm{{\sc
a}}}D_1+ d_2 \, \phantom{}_{\textrm{{\sc a}}}D_2$, by setting
$d_2=0$.\\
Now when this solution was allowed evolve to the threshold,
$D|_{x=0}=d_3 \,\phantom{}_{\textrm{{\sc t}}}D_1+ d_4 \,
\phantom{}_{\textrm{{\sc t}}}D_2$, the constants $d_3, d_4 \neq 0$
were fixed [see (\ref{dconstants})]. To match across $x=0$, we
must have (since we are using a global coordinate system) \be
D^M|_{x=0}=D^V|_{x=0}, \ee where $M$ denotes solutions coming from
Minkowski space-time and $V$ denotes solutions coming from Vaidya
space-time. Thus we require \be \lim_{x\uparrow 0} ( d_3 \,
\phantom{}_{\textrm{{\sc t}}}D_1^M+d_4 \, \phantom{}_{\textrm{{\sc
t}}}D_2^M )=\lim_{x\downarrow 0} ( d_5 \, \phantom{}_{\textrm{{\sc
t}}}D_1^V+d_6 \, \phantom{}_{\textrm{{\sc t}}}D_2^V ). \ee From
(\ref{nninZhyp}),(\ref{ninZ}), we see the solutions from Minkowski
space-time are $O(1),O(x^{s+2})$. When $s \in \mathbb{Z} \geq 2$,
we see from (\ref{dsolsint}) the solutions from Vaidya space-time
are also $O(1),O(x^{s+2})$, and thus to match continuously across
$x=0$ we cannot switch off the $O(1)$ $D$-solution.\\
Thus when we calculate $A$, and hence $\delta \Psi_4$, there will
be divergence as $x \downarrow 0$ due to this solution. This does
not happen when $s \notin \mathbb{Z}$, since there is no divergent
$A$-solution when $Re(s) >2$.\\
There were four constraints for initial data: (i) finite flux on
the axis, (ii) finite flux on the threshold when approached from
flat space, (iii) finite flux on the threshold when approached
from Vaidya space-time, and (iv) continuous matching across $x=0$.
The most general class of perturbations which satisfies all of
these conditions are those with \be Re(s)>2, \quad s \notin
\mathbb{Z}. \ee

\subsubsection{Cauchy Horizon.}

When $\lambda<1/8$, the Cauchy horizon is a regular singular point
of the system given in Appendix B. Its leading order coefficient
matrix has eigenvalues $0$ multiplicity three and \be \sigma
\equiv \f{1}{2}\left(s-4+\f{s}{\sqrt{1-8\lambda}}\right). \ee When
$\sigma$ is not an integer, we can use the system methods outlined
in Appendix C. Applying Theorem \ref{theorem:thone} we find
solutions with asymptotic behaviour \be
\begin{array}{cc|c|c|c} &
I & II & III & IV \\
A= & O(1) & O(w^{\sigma})& O(1) & O(1) \\
D= & O(1) & O(w^{\sigma})& 0 & 0 \\
K= & O(w) & O(w^{\sigma+1})& 0 & O(1) \\
G= & O(w) & O(w^{\sigma+1}) & O(1) & 0
\end{array} \label{tablexc}
\ee as $w \rightarrow 0$ where $w=x-x_c$ (for consistency see (\ref{psymbol})).\\
Now we make an important observation: since \be 0<\sqrt{1-8
\lambda}< 1 \ee for $0 < \lambda <1/8$, therefore \be
\sigma=\f{1}{2}\left(s-4+\f{s}{\sqrt{1-8\lambda}}\right)
> \f{1}{2}(2s-4), \ee and thus $Re(\sigma) > 0$ for $Re(s)> 2$. Alternatively,
we can say \be \sigma=s-2+ O(\lambda) \ee where each coefficient
of $\lambda^n$ is positive, and thus again $Re(\sigma) > 0$ for
$Re(s)> 2$. Thus each solution for $A$ and $D$ as given in
(\ref{tablexc}) is at most $O(1)$ near $x=x_c$; all the solutions
for $A$ and $D$ which are series beginning with $w, w^\sigma$ or
$w^{\sigma+1}$ will decrease to zero as we approach the Cauchy
horizon.

Since \be \delta \Psi_0=\f{2r^{s-3}D}{\lambda x-1}, \quad \delta
\Psi_4=\f{r^{s-3}(2A+(1- \lambda x)D)}{2},\ee and $A$ and $D$ near
the Cauchy horizon are a linear combination of $O(1)$ solutions,
the scalars $\delta \Psi_{0,4}$ representing the flux of the
perturbation, and hence the scalar $\delta P_{-1}$, will be finite
on the Cauchy horizon $x=x_c$. Thus when $\sigma \notin
\mathbb{Z}$, the Cauchy horizon
is stable under metric and matter perturbations.\\

However, for each value of the parameter $\lambda <1/8$, there
will be a mode number $s$ such that $\sigma \in \mathbb{Z}$, and
thus we must also consider this case. From (\ref{psymbol}), we see
a general solution for $D$ near $w=x-x_c=0$ can be written as \be
D&=& d_3 \sum_{m=0}^{\infty} A_m w^{m+\sigma}+d_4 \left\{ k \ln w
\sum_{m=0}^{\infty} A_m w^{m+\sigma}+\sum_{m=0}^{\infty} B_m
w^{m}\right\}, \ee where $d_3,d_4$ are constants and $k$ can be
zero. Since we are considering $\lambda >0$ ($\lambda=0$ being
vacuum space-time) and $Re(s)> 2$, we have $\sigma \geq 1$ if
$\sigma \in \mathbb{Z}$. Now we use this solution for $D$ as an
inhomogeneous term to integrate the perturbation equations. Near
$w=0$, we find a 4-parameter set of solutions: \be K&\sim &k_0+
\f{x_c^{-1}(x_c-x_e)}{1-\lambda x_c} \int w D'
dw-x_c^{-1}(x_c+3-n) \int D dw, \nonumber \\
G&\sim&g_0+\f{1}{x_c(\lambda x_c-1)}\int D dw, \nonumber \\
A&\sim& \f{1}{x_c(1-\lambda x_c)}D+2\lambda G+\tfrac{1}{2}
x_c^{-1}[\lambda x_c-x_c(l^2+l-2)+6] K- \tfrac{1}{2}(\lambda
x_c^2+4) K',\ee where $k_0,g_0$ are constants, a prime denotes
differentiation w.r.t. $w$, and \be \int D dw&=& d_3
\sum_{m=0}^{\infty} \f{A_m w^{m+\sigma+1}}{m+\sigma+1}+d_4 \left\{
k \sum_{m=0}^{\infty} \f{A_m w^{m+\sigma+1}}{m+\sigma+1}\left(\ln
w-\f{1}{m+\sigma+1}\right)+\sum_{m=0}^{\infty} \f{B_m
w^{m+1}}{m+1}\right\} , \nonumber \\ \int w D' dw&=&d_3
\sum_{m=0}^{\infty} \f{A_m (m+\sigma)
w^{m+\sigma+1}}{m+\sigma+1}+d_4 \left\{ k \sum_{m=0}^{\infty}
\f{A_m w^{m+\sigma+1}}{m+\sigma+1}\left(1+ (m+\sigma)\ln
w-\f{m+\sigma}{m+\sigma+1}\right)+\sum_{m=0}^{\infty} \f{B_m m
w^{m+1}}{m+1}\right\}. \nonumber \ee Since $\sigma \geq 1 $ and
$\lim_{w\rightarrow 0} w^\sigma \ln w =0$, we see all of these
variables $A,D,K,G,$ and thus the scalars $\delta \Psi_{0,4}$ and
$\delta P_{-1}$, are again finite in the limit $w\rightarrow 0$.

Thus the full set of perturbations which are finite on the axis
and on the threshold $\mathcal{N}$ will evolve up to the Cauchy
horizon and beyond \emph{without} their flux diverging. Therefore
in the case of self-similar null dust there is a naked singularity
whose Cauchy horizon is stable under metric and matter
perturbations.

\subsubsection{Second Future Similarity Horizon.}

Now something interesting happens when we allow the solution to
evolve past the Cauchy horizon and on to the next singular
surface, the SFSH given by
$x_e=\tfrac{1}{2\lambda}(1+\sqrt{1-8\lambda})$. The first scalar
depends only on $D$, \be \delta\Psi_0=\f{2 r^{s-3} }{\lambda
x-1}D,\ee and the solutions for $D$ near $x=x_e$ can be found
directly from (\ref{psymbol}) as \be D=d_1\sum_{m=0}^{\infty} A_m
(x-x_e)^m +d_2 \sum_{m=0}^{\infty} B_m (x-x_e)^{m+\varsigma}, \ee
where \be
\varsigma=\f{1}{2}\left(s-4-\f{s}{\sqrt{1-8\lambda}}\right).\ee
Since $0<\sqrt{1-8\lambda}<1,$ we see $\varsigma$ will always be
negative for $Re(s) \geq 2$. Thus there is a class of solutions
which are finite on the axis, finite on the threshold
$\mathcal{N}$, finite on the Cauchy horizon, and then finally
diverge on the SFSH. We emphasize that this instability is due to
$x=x_e$ being a similarity horizon of the space-time, and not an
event horizon.

\subsection{$l=1$ mode, Vaidya.}

In this sector we can only define partially gauge invariant
objects. As in Minkowski space-time, Section III.B, the metric
perturbation objects are gauge sensitive to $\xi_\mu dx^\mu=\xi_A
Y dx^A+ \xi Y,_a dx^a$ as \be k_{AB}& \rightarrow &
k_{AB}+\left[r^2(\xi/r^2),_A\right]_{|B}+\left[r^2(\xi/r^2),_B\right]_{|A}\\
k& \rightarrow & k+2\xi/r^2+2\w{v}^A r^2 (\xi/r^2),_A. \ee We
examine the case where the perturbed space-time is that of null
dust, and thus the bare matter perturbations are as given in
(\ref{mattpert}),
\be \Delta t_{AB}=\left(%
\begin{array}{cc}
  r^2 \delta\rho-\f{\lambda\partial_x\Gamma}{4\pi r} & r x \delta\rho-\f{\lambda\partial_r\Gamma}{8\pi r}-\f{\lambda x \partial_x\Gamma}{8\pi r^2} \\
  \textrm{Symm} & x^2 \delta\rho-\f{\lambda x \partial_r\Gamma}{4\pi r^2} \\
\end{array}%
\right), \qquad
\Delta t_A=\left(%
\begin{array}{c}
  \f{-\lambda \Gamma}{8\pi r} \\
  \f{-\lambda x \Gamma}{8\pi r^2} \\
\end{array}%
\right), \qquad \Delta t^1=\Delta t^2=0.  \ee Since $p_A=h_A$ in
this sector, we can transform into the equivalent of the
Regge-Wheeler gauge by choosing \be \xi_A=h_A-r^2(\xi/r^2)_{|A},
\ee the benefit being that in this gauge $p_A=0$ and therefore
$T_{AB}=\Delta t_{AB}$ etc. Thus we can express the right hand
side of the perturbation equations given in Appendix A in terms of
$\delta
\rho, \Gamma$.\\
Further transformations maintain this condition provided
$\xi_A=-r^2(\xi/r^2)_{|A}$. Importantly, this fixes $\xi_A$ while
keeping $\xi$ completely free.\\
$T_{AB}$ and $T_A$ are not gauge invariant, they are sensitive to
gauge transformations as  \be \ov{T}_{AB}-T_{AB} & = &
\w{t}_{AB|C}r^2(\xi/r^2)^{|C} +\w{t}_{CB}(r^2
(\xi/r^2)^{|C})_{|A}+\w{t}_{CA}(r^2(\xi/r^2)^{|C})_{|B},
\nonumber \\
  \ov{T}_A- T_A & = & -\w{t}_{AB}r^2(\xi/r^2)^{|B}.
\ee The vanishing of $T^1$ and $T^2$ is gauge invariant.\\
Now we look at the perturbation equations in $(x,r)$ coordinates.
As in the $l\geq 2$ sector we perform a Mellin transformation of
the equations, which is equivalent to parameterizing the
perturbation components as in (\ref{mellofpert}),
\be k_{AB}=\left(%
\begin{array}{cc}
  r^{s+1}A_s(x) & r^s B_s(x) \\
  r^s B_s(x) & r^{s-1} C_s(x) \\
\end{array}%
\right), \qquad k=r^{s-1} K_s(x), \qquad \Gamma=r^s G_s(x), \qquad
\delta \rho=r^{s-3} M_s(x). \ee Again, dropping the subscript $s$,
we define the new variable $D=B-x A$.\\
We exploit the gauge freedom and transform into a gauge in which
$k_A^{\ph{A}A}=0$, by choosing $\xi$ such that \be
\left[r^2(\xi/r^2),_A\right]^{|A}=-k_A^{\ph{A}A}.\ee Then we are
allowed make further transformations which preserve RW gauge and
$k_A^{\ph{A}A}=0$ provided \be
\left[r^2(\xi/r^2),_A\right]^{|A}&=&0, \nonumber \\
\xi_A+r^2(\xi/r^2)_{|A}&=&0.\ee Thus we recover the perturbation
equation (\ref{tracek}), which was not valid in the $l=1$ sector,
as a gauge choice. As before the set of perturbation equations in
this gauge reduces to one constraint equation defining $\delta
\rho$, a first order system in $A,D,K$ and $G$,
and we can decouple a second order ordinary differential equation for $D$.\\

There is some gauge freedom in $\xi$ left, and we will use this to
gauge away $k$. Let us formalise this: let $\xi$ satisfy gauge
conditions $L_1 \xi=0$; let $k$ satisfy perturbation field
equation $L_2 k=0$; and let $k$ transform as $k \rightarrow k+L_3
\xi$. Since $L_2 L_3 \xi =0$ subject to $L_1 \xi=0$, there is a
gauge in which $k=0$.\\
In the coordinate system $(x,r)$, where we perform a Mellin
transform over $r$ such that $\xi=r^{s-1} \xi_s$, the above
equations are \be L_1 \xi&=& - \left( -1 + s \right) \,\left( -
x\,\lambda    + s\,\left( -1 + x\,\lambda \right)
         \right) \,\xi_s (x)   + \left( -2 + 2\,x - 3\,x^2\,\lambda  + 2\,s\,\left( 1 - x + x^2\,\lambda  \right)  \right) \,
     \xi_s '(x) \nonumber \\ && \qquad  + x\,\left( -2 + x - x^2\,\lambda  \right) \,\xi_s ''(x)
     , \nonumber \\
     L_2 \xi &=& \xi_s'''(x) +p_1(x)\, \xi_s''(x)+q_1(x)\,
\xi_s'(x)+r_1(x)\, \xi_s(x), \nonumber \\ L_3 \xi &=&  \left( s +
x\,\lambda - s\,x\,\lambda  \right) \,\xi_s (x) +
    \left( 1 - x + x^2\,\lambda  \right) \,\xi_s '(x),  \ee

where

\begin{small}
\be p_1(x)&=&\frac{6 + 5\,\left( -2 + x \right) \,x - 4\,x\,\left(
-3 + x + 3\,x^2 \right) \,\lambda  +
    x^3\,\left( 24 + 7\,x \right) \,{\lambda }^2 + n^2\,\left( -1 + x\,\lambda  \right) \,
     \left( 4 + 3\,x\,\left( -1 + x\,\lambda  \right)  \right)}{x\,\left( 1 + n - x + x\,\left( 3 - n + x \right) \,\lambda  \right) \,
    \left( 2 + x\,\left( -1 + x\,\lambda  \right)  \right)
    }\nonumber \\ && +\frac{
    n\,\left( 2 + x\,\left( 1 - 16\,\lambda  + x\,\left( -3 +
             \lambda \,\left( 16 + x\,\left( 6 - \left( 17 + 3\,x \right) \,\lambda  \right)  \right)  \right)  \right)  \right) }
    {x\,\left( 1 + n - x + x\,\left( 3 - n + x \right) \,\lambda  \right) \,
    \left( 2 + x\,\left( -1 + x\,\lambda  \right)  \right)
    },\nonumber \\
q_1(x)&=&\frac{\left( 2 - n \right)\left( -2n\left( 1 + n \right)
+ \left( 1 + n \right)\left( -1 + 3n \right) x +
       \left( 1 - 3n \right)x^2 \right)  - 2x
     \left( 3 + n\,\left( -1 + \left( -3 + n \right) \,n \right)  - x - \left( -3 + n \right) \,n\,\left( -4 + 3\,n \right) \,x\right)}{x^2\,
    \left( 1 + n - x + x\,\left( 3 - n + x \right) \,\lambda  \right) \,\left( 2 + x\,\left( -1 + x\,\lambda  \right)  \right) } \nonumber \\ &&
    +\frac{
       \left(-2x\left( 5 + 3\,\left( -3 + n \right) \,n \right) \,x^2 \right) \,\lambda  +
    \left( -1 + n \right) \,x^3\,\left( -30 - 8\,x + n\,\left( 19 - 3\,n + 3\,x \right)  \right) \,{\lambda }^2}{x^2\,
    \left( 1 + n - x + x\,\left( 3 - n + x \right) \,\lambda  \right) \,\left( 2 + x\,\left( -1 + x\,\lambda  \right)  \right) }
    ,\nonumber \\
    r_1(x)&=&- \frac{\left( -1 + n \right) \,\left( \left( 2 - n \right) \,\left( 1 + n \right) \,\left( n - x \right)  +
        2\,\left( 3 + n^3\,x + x\,\left( 5 + x \right)  + 2\,n\,\left( 1 + x^2 \right)  -
           n^2\,\left( 1 + x\,\left( 4 + x \right)  \right)  \right) \,\lambda\right) }{x^2\,
      \left( 1 + n - x + x\,\left( 3 - n + x \right) \,\lambda  \right) \,\left( 2 + x\,\left( -1 + x\,\lambda  \right)  \right) }
      \nonumber
      \\&&+\frac{(-1+n)\left(
        \left( -3 + n \right) \,n\,\left( -4 + n - x \right) \,x^2\,{\lambda }^2 \right) }{x^2\,
      \left( 1 + n - x + x\,\left( 3 - n + x \right) \,\lambda  \right) \,\left( 2 + x\,\left( -1 + x\,\lambda  \right)  \right) }
   . \nonumber
\ee

\end{small}
A direct consequence of $k=0$ is that $D=0$. Thus we are in a
gauge in which $p_A=k_A^{\ph{A}A}=k=D=0$, and to remain in this
gauge we are allowed further transformations provided \be
\left[r^2(\xi/r^2),_A\right]^{|A}&=&0, \nonumber \\
2\xi/r^2+2\w{v}^A r^2 (\xi/r^2),_A&=&0. \ee The only $\xi$ which
satisfies both these constraints is $\xi=0$. Therefore there is no
remaining gauge freedom and so the remaining perturbation variables are gauge invariant.\\
Thus we have found a one parameter family of solutions, \be && k_{AB}=\left(%
\begin{array}{cc}
  2 g_0 \lambda r^{s+1} x^{s} & 2 g_0 \lambda r^s x^{s+1} \\
  2 g_0 \lambda r^s x^{s+1} & 2 g_0 \lambda r^{s-1} x^{s+2} \\
\end{array}%
\right), \quad T_{AB}=\left(%
\begin{array}{cc}
  \f{\lambda g_0 (s+x) r^{s-1} x^{s-1}}{4 \pi} & \f{\lambda g_0 (s+x) r^{s-2} x^{s}}{4 \pi} \\
  \f{\lambda g_0 (s+x) r^{s-2} x^{s}}{4 \pi} & \f{\lambda g_0 (s+x) r^{s-3} x^{s+1}}{4 \pi} \\
\end{array}%
\right),\nonumber \\&& T_A=\left(%
\begin{array}{c}
  \f{-\lambda g_0 r^{s-1} x^{s}}{8 \pi} \\
  \f{-\lambda g_0 r^{s-2} x^{s+1}}{8 \pi} \\
\end{array}%
\right), \quad T^1=T^2=k=0. \ee To match continuously with the
empty Minkowski region preceding $x=0$ we require $Re(s)\geq 1$.
These solutions will evolve without divergence through the rest of
the space-time.\\
Note the perturbations will vanish if $g_0=0$, that is to say if
we had considered only metric perturbations, and no matter
perturbations, we would have returned an empty sector. Note also
the sector is empty in the Minkowski limit $\lambda \rightarrow
0$.

\subsection{$l=0$ mode, Vaidya.}

When $l=0$, then $Y,_a=0$ and our perturbations are \be h_{\mu
\nu}=\left(
\begin{tabular}{cc} $h_{AB}$ & 0 \\  0
& $r^2 K \gamma_{ab}$ \end{tabular} \right), \qquad \Delta
t_{\mu\nu}=\left(
\begin{tabular}{cc} $\Delta t_{AB}$ & 0 \\  0
& $r^2 \Delta t^1 \gamma_{ab}$ \end{tabular} \right). \ee We will
use the coordinates $(v,r,\theta,\phi)$, where $v$ is the null
coordinate of the background. Thus the metric of the background is
$\w{g}_{\mu\nu}dx^\mu dx^\nu=-(1-\frac{\lambda v}{r})dv^2 + 2 dv\,
dr+r^2 \gamma_{ab}dx^a dx^b$.\\
Since the matter tensor has the form (\ref{pertnulldust}) and
$\w{\ell}_{\mu}$ has no angular dependence, we find $\Delta
t^1=0$. We can describe the ingoing radial null geodesic of the
perturbed space-time as $\w{\ell}_\mu+\delta \ell_\mu=-\nabla_\mu
V$ where $V=v+\Gamma(v,r)$ is the null coordinate of the perturbed
space-time. Our unknowns
therefore are $h_{AB}, K, \delta \rho$ and $\Gamma$.\\
We cannot construct gauge invariants in the $l=0$ sector and thus
we exploit the remaining gauge freedom to set some variables equal
to zero. The perturbation variables are gauge dependent as \be
h_{AB} &\rightarrow& h_{AB}-(\xi_{A|B}+\xi_{B|A}), \nonumber \\
K &\rightarrow& K-2 \w{v}^A \xi_A, \nonumber \\
\Delta t_{AB} &\rightarrow& \Delta
t_{AB}-\w{t}_{AB|C}\xi^C-\w{t}_{CB}\xi^C_{\phantom{C}|A}-\w{t}_{CA}\xi^C_{\phantom{C}|B}.
\label{gtlzv}\ee As before, we transform into a gauge in which
$K=h_{\ph{A}A}^A=0$. To do this we choose $\xi_A$ such that (in
the $(v,r)$ coordinate system where $\dot{\theta}$ and $\theta'$
denote differentiation w.r.t. $v$ and $r$ respectively) \be \xi_v
+(1-\tfrac{\lambda v}{r})\xi_r&=&\tfrac{1}{2} r K, \nonumber \\
\dot{\xi_r}&=& \tfrac{1}{2}h_{\ph{A}A}^A. \ee Then we are free to
make further gauge transformations which preserve this condition
provided \be \xi_v +(1-\tfrac{\lambda v}{r})\xi_r&=&0, \nonumber
\\ \dot{\xi_r}&=& 0. \label{rglzv}\ee Now we look at the field equations in this gauge (we will
let $h_{vv}=A, \, h_{vr}=h_{rv}=B$ and $h_{rr}=C$). The first is
\be \dot{C}=0, \ee and thus $C=C(r)$. When we perform a gauge
transformation on this quantity subject to (\ref{gtlzv}) and
(\ref{rglzv}), we find $C \rightarrow C - 2 \xi_r'$, but since
$\xi_r$ is an arbitrary function of $r$, this means we can choose
a gauge in which $C=0$ (and thus $B=0$ since $h_{\ph{A}A}^A=0$).
Thus we have transformed to a gauge in which
$K=h_{\ph{A}A}^A=B=C=0$, and to remain in this gauge we are
allowed further gauge transformations of the form
$\xi_A=c_0(-(1-\tfrac{\lambda v}{r})
\delta_A^{\,v}+\delta_A^{\,r})$, with $c_0$ an arbitrary
constant.\\ The remaining perturbation equations in this gauge are
\be r A''+2 A' &=&0, \nonumber \\ r A'+A+ \lambda \Gamma'&=&0,
\nonumber \\ \tfrac{1}{r^2}(1-\tfrac{\lambda v}{r})(r
A'+A)+\tfrac{1}{r} \dot{A}-\tfrac{2 \lambda}{r^2} \dot{\Gamma}&=&8
\pi \delta \rho.\ee That $\ell_\mu+\delta \ell_\mu$ must be null
and geodesic gives $\Gamma'=0$, and hence \be
\Gamma=\alpha(v),\qquad A=\f{\beta(v)}{r}, \qquad \delta \rho
=\f{1}{8 \pi r^2}(\dot{\beta}-2 \lambda \dot{\alpha}). \ee Further
gauge transformations give \be A &\rightarrow& A-2
\tfrac{\lambda}{r} c_0, \nonumber \\
\Delta t_{AB} &\rightarrow& \Delta t_{AB}, \ee and thus these
remaining perturbation quantities cannot be gauged away. \\
What we have shown here is essentially a uniqueness result: all
the above perturbations can be generated by a perturbation in the
mass function and the null vector. The metric and matter tensors
for spherically symmetric null dust are given by \be
g_{\mu\nu}dx^\mu dx^\nu=-\left(1-\f{m(v)}{r}\right)dv^2+2dv dr+r^2
d\Omega^2, \nonumber
\\ t_{\mu\nu}dx^\mu dx^\nu=\frac{\dot{m}(v)}{8 \pi r^2} \ell_\mu \ell_\nu,
\ee where before perturbation $m(v)=\lambda v$ and
$\ell_\mu=-\p_\mu v$, and after a perturbation $m(v)=\lambda v
+\beta(v)-2 \lambda c_0$ and $\ell_\mu=-\p_\mu(v+\alpha(v))$. The
terms $\alpha,\beta, c_0$ are arbitrary and therefore there is no
reason to suspect divergence on the Cauchy horizon or elsewhere.

\section{Resummation}
In this section, we review some properties of the Mellin transform
(see e.g.\ \cite{davies}, \cite{henrici}) and present a
plausibility argument that finiteness of the modes (as seen above
at the Cauchy horizon) implies finiteness of the full perturbation
found by resumming the modes. This argument is based on a study of
the Klein-Gordon equation (or wave equation) in Vaidya space-time,
and its correspondence to the wave equation in Minkowski
space-time.

We recall that the line element of self-similar Vaidya space-time
may be written as \be ds^2 = r^2(-1+\lambda x)dx^2+2r(1-x+\lambda
x^2)dx\,dr+x(2-x+\lambda
x^2)dr^2+r^2d\Omega^2,\label{vaidyalel}\ee with the coordinate
ranges $r\in [0,\infty), x\in(-\infty,\infty)$. We note that
Minkowski space-time corresponds to taking $\lambda =0$. Then the
wave equation (or more accurately, the PDE satisfied by the
$l^{\rm th}$ multipole moment $\phi=\phi_l$ of the Klein-Gordon
field) is \be -x(2-x+\lambda x^2)\phi_{,xx}+2(1-x+\lambda
x^2)r\phi_{,rx}+(1-\lambda x)r^2\phi_{,rr}-\lambda
x^2\phi_{,x}+(2-\lambda x)r\phi_{,r}-l(l+1)\phi=0.\label{waveq}\ee
The Mellin transform of this equation yields a parametrised ODE.
The first step is to define the Mellin transform of the field
$\phi$ (note that the arrow indicates the variable with respect to
which the Mellin transform is taken) : \be P(x;t)=
{\cal{M}}[\phi(x,r)](r\to t):= \int_0^\infty \phi(x,r)r^{t-1}\,dr,
\label{mellin}\ee where $t\in \mathbb{C}$ and satisfies
$\tau_1\leq\mathrm{Re}(t)\leq\tau_2$, the constants $\tau_{1,2}$
being defined by the condition that the integral converges for
these values of $t$. The field $\phi(x,r)$ is recovered via the
inverse Mellin transform: \be
\phi(x,r)={\cal{M}}^{-1}[P(x;t)](t\to r)=\frac{1}{2\pi
i}\int_{\tau-i\infty}^{\tau+i\infty}P(x;t)r^{-t}\,dt,\label{invmell}\ee
where the inversion contour is vertical and lies in the strip
$\tau_1\leq\mathrm{Re}(t)\leq\tau_2$. We should note here that as
the resummation is over terms of the form $P(x;t)r^{-t}$, the role
played by $t$ in the present section corresponds to the role
played by $-s$ in Sections 2 - 5. See in particular the last
paragraph of Section 2.

Integrating by parts immediately yields the following results:
\begin{lemma}
\be {\cal{M}}[r\phi_{,r}(x,r)](r\to t)=-t{\cal{M}}[\phi(x,r)](r\to
t)=-tP(x;t),\ee provided \be\lim_{r\to
0}r^t\phi(x,r)=\lim_{r\to\infty}r^t\phi(x,r)=0.\label{melcon1}\ee
\end{lemma}
\begin{lemma}
\be {\cal{M}}[r^2\phi_{,rr}(x,r)](r\to
t)=t(t+1){\cal{M}}[\phi(x,r)](r\to t)=t(t+1)P(x;t),\ee provided
(\ref{melcon1}) holds and \be\lim_{r\to
0}r^{t+1}\phi_{,r}(x,r)=\lim_{r\to\infty}r^{t+1}\phi_{,r}(x,r)=0.\label{melcon2}\ee
\end{lemma}

Assuming that these conditions hold for values of $t$ on a strip
of the complex plane, we can take the Mellin transform of
(\ref{waveq}) to obtain the ODE \be -x(2-x+\lambda
x^2)P^{\prime\prime}-(2t(1-x)+(1+2t)\lambda
x^2)P^\prime+(t^2-t-l(l+1)-t^2\lambda x)P=0,\label{melltreq}\ee
where the prime here and throughout refers to differentiation with
respect to argument. This equation has analytic coefficients (and
hence analytic solutions) everywhere except at infinity and at the
singular points $x=0,x_c,x_e$, the roots (in increasing order) of
$x(2-x+\lambda x^2)=0$. Note that in Minkowski space-time, there
is no third singular point $x_e$. These are all regular singular
points of the equation, and thus the standard Frobenius theory can
be used to study the global behaviour of solutions of
(\ref{melltreq}) (see for example \cite{CoddLev}, or any textbook
on linear differential equations). We recall that $x=x_0=$constant
is a space-like hypersurface for $x_0\in(0,x_c)$ and that
$x=0,x_c$ are null hypersurfaces.

For Minkowski space-time, $\lambda=0$ and (\ref{melltreq}) is a
hypergeometric differential equation, and we can give an
essentially complete account of the problem at hand. We proceed to
do so in order to clarify the nature of this problem and our
putative solution. So let $\lambda=0$ in (\ref{melltreq}) and for
convenience of comparison with the standard text by Bateman
\cite{Bate}, let $x=2z$ (all results quoted below are taken from
this reference). Then (\ref{melltreq}) is the hypergeometric
equation \be
z(1-z)u^{\prime\prime}+(c-(a+b+1)z)u^\prime-abu=0,\label{hyper}\ee
where $u(z)=P(x), a=t+l, b=t-l-1$ and $c=t$. We will assume that
$t\not\in \mathbb{Z}$. This can be assumed without loss of
generality by a deformation of the inversion contour.

The past and future null cones of the origin then correspond to
the singular points $z=0,1$ respectively of this equation. We
encounter here a slight difficulty. Being a null hypersurface,
$x=z=0$ cannot be an initial data surface for the equation
(\ref{waveq}). We expect that this will translate into $z=0$
failing to be a `good' initial point for the ODE (\ref{hyper}).
However, our overall aim is to argue that finiteness of the field
on $x=0$ along with finiteness of the modes at the future null
cone (Cauchy horizon) is sufficient to imply finiteness of the
field at the future null cone. We can connect these two by
determining their respective connections to Cauchy data on a
space-like hypersurface, for example, $x=1 (z=1/2)$. So consider
the Cauchy data
\[ \alpha(r)=\phi|_{x=1},\quad \frac{\beta(r)}{2}=\phi_{,x}|_{x=1}.\]
We assume that $\alpha, \beta$ satisfy unspecified
differentiability and integrability conditions that, in
particular, allow us to calculate the Mellin transforms
\[ a(t)={\cal{M}}[\alpha(r)](r\to t),\quad
b(t)={\cal{M}}[\beta(r)](r\to t)\] on some strip $\tau_1\leq
\mathrm{Re}(t)\leq\tau_2$ of the complex plane (the factor 1/2 is
given for later convenience). These then yield the initial data
for the Mellin transform $P(x;t)$ of $\phi$ at $x=1$, i.e.\ for
$u(z)$ at $z=1/2$:
\[ u(\frac12)=a(t),\quad u^\prime(\frac12)=b(t).\]
We emphasise that our assumptions on $\alpha,\beta$ imply the
existence of the inverse Mellin transform of $a,b$ taken over a
vertical contour in $\tau_1\leq \mathrm{Re}(t)\leq\tau_2$.

The next step is to determine the solution for $u$ at $z=0$ and at
$z=1$ in terms of $a(t)$ and $b(t)$. We use the following pairs of
linearly independent solutions at these two points (again
following the notation of Bateman \cite{Bate}). At $z=0$ we use
\[ u_1(z;t)=(1-z)^{1-t}F(-l,l+1;t;z),\quad
u_5(z;t)=z^{1-t}F(l+1,-l;2-t;z),\] and at $z=1$ we use
\[ u_2(z;t)=z^{1-t}F(l+1,-l;t;1-z),\quad
u_6(z;t)=(1-z)^{1-t}F(-l,l+1;2-t;1-z),\] where \[F(a,b;c;z)=
{}_2F_1(a,b;c;z)=\sum_{n=0}^\infty\frac{(a)_n(b)_n}{n!(c)_n}z^n,\]
with \[(a)_0=1,\quad (a)_n=a(a+1)\cdots(a+n-1),\quad
n=1,2,3,\dots\] is the standard hypergeometric function. Note then
that all the hypergeometric functions present are in fact
polynomials, and so $u_5$ and $u_6$ are finite sums of powers of
$z$ and $1-z$ respectively. $u_1$ and $u_2$ are infinite series in
$z$ and $1-z$ respectively, both with radius of convergence at
least 1. We will concentrate on the solution at $z=0$. The general
solution of (\ref{hyper}) at $z=0$ is \be
u(z;t)=c_1(t)u_1(z;t)+c_5(t)u_5(z;t),\label{tolat0}\ee giving \be
u^\prime(z;t)=c_1(t)u_1^\prime(z;t)+c_5(t)u_5^\prime(z;t).\label{dsolat0}\ee
Note that existence of $u_5|_{z=0}$ and $u_5^\prime|_{z=0}$
requires $\mathrm{Re}(t)\leq 0$: ruling out the integer case, we
take $\mathrm{Re}(t)<0$. As $z=1/2$ is within the radius of
convergence of the series solutions, we have
\begin{eqnarray*}
c_1(t)u_1(\frac12;t)+c_5(t)u_2(\frac12;t)&=&a(t),\\
c_1(t)u_1^\prime(\frac12;t)+c_5(t)u_2^\prime(\frac12;t)&=&b(t).
\end{eqnarray*}
Solving for $c_1(t),c_2(t)$ gives
\begin{eqnarray*}
c_1(t)&=&\frac{2^{1-2t}}{1-t}(a(t)u_5^\prime(\frac12;t)-b(t)u_5(\frac12;t)),\\
c_5(t)&=&\frac{2^{1-2t}}{1-t}(-a(t)u_1^\prime(\frac12;t)+b(t)u_1(\frac12;t)),
\end{eqnarray*}
where we have used Abel's formula for the Wronskian which arises
as a determinant in solving for $c_1,c_5$. We can obtain similar
expressions for $c_2(t),c_6(t)$, the coefficients of
$u_2(z;t),u_6(z;t)$ in the general solution for $u$ at $z=1$.

Our next step is to address the following question. Do the
conditions on $a(t),b(t)$ (i.e.\ existence of the inverse Mellin
transform in the strip $\tau_1\leq \mathrm{Re}(t)\leq \tau_2$)
imply the existence of the inverse Mellin transforms of \bes
P(0;t)&=&c_1(t)u_1(0;t)+c_5(t)u_5(0,t),\\
P^\prime(0;t)&=&\frac12(c_1(t)u_1^\prime(0;t)+c_5(t)u_5^\prime(0,t)),\ees
and of the corresponding expressions at $x=1$? By linearity, this
is true if and only if it is separately true for $a$ and for $b$.
We consider only the case $a\neq 0$, $b=0$: the reverse case is
similar. Then noting that $u_1(0;t)=1$ and that
$u_5(0;t)=u_5^\prime(0;t)=0$ (since $\mathrm{Re}(t)<0$), we have
\bes t_0(t)&:=&
P(0;t)=\frac{2^{1-2t}}{1-t}u_5^\prime(\frac12;t)a(t)\\
&=& \sum_{k=0}^\infty
c_{k,l}\frac{2^{-2t}}{(t-1)(t-2)\cdots(t-k)}a(t)\ees where
\[c_{k,l}=(-1)^k\frac{2}{k!}(l+1)_k(-l)_k,\quad k=0,1,\dots\]
The rational function of $t$ can be written as the sum of inverse
linear terms using
\[
\frac{1}{(t-1)(t-2)\cdots(t-k)}=\sum_{j=1}^k\frac{(-1)^{k-j}}{(k-j)!(j-1)!}\frac{1}{t-j}.\]
We now point out two properties of the Mellin transform that will
allow us to perform the required inversion.

\begin{lemma}
If \[ G(t)={\cal{M}}[\gamma(r)](r\to t),\] then
\[ k^tG(t)={\cal{M}}[\gamma(\frac{r}{k})](r\to t)\]
for $0<k\in \mathbb{R}$, provided all relevant integrals converge.
\end{lemma}
\noindent{\bf Proof:} This is immediate from the formula
(\ref{invmell}) for the inverse Mellin transform.

\begin{lemma}
If \[ G(t)={\cal{M}}[\gamma(r)](r\to t),\] then
\[\frac{G(t)}{k+t}={\cal{M}}[-r^k\int_0^r\frac{\gamma(y)}{y^{k+1}}\,dy](r\to
t)\] for $k\in \mathbb{R}$, provided all relevant integrals
converge. \end{lemma} \noindent{\bf Proof:} This follows from
Lemma 6.1.

Using these results, we could give a closed form expression for
the inverse Mellin transform of $t_0(t)$ in terms of integrals of
$\alpha(r)$, the inverse Mellin transform of $a(t)$. These will
have the structure of a (terminating) series, whose $k^{th}$ term
consists of a sum of $k$ terms, the $j^{th}$ of which has the form
\be
d_{j,k}(\frac{r}{4})^{-j}\int_0^{r/4}\alpha(y)y^{j-1}\,dy\label{crucialintt}\ee
for some constants $d_{j,k}$. Note that the $r/4$ arises from the
term $2^{-2t}$ in $t_0(t)$ and by using Lemma 6.3. Since there is
only a finite number of terms present, the only questions
regarding convergence relate to the existence of the integrals
(\ref{crucialintt}). This existence follows from the conditions
laid down on the initial data function $\alpha(r)$. In exactly the
same way, we can deduce the existence of the inverse Mellin
transform of
\[ t_1(t)=P(2;t)=u(1;t)=\frac{2^{1-2t}}{t-1}u_6^\prime(\frac12;t)a(t),\]
and of the terms corresponding to $t_0(t)$ and $t_1(t)$ that arise
by taking $a(t)=0, b(t)\neq 0$.

To conclude the discussion for Minkowski space-time, we take a
slightly different point of view, where we treat
$u(0;t)=c_1(t)u_1(0;t)+c_5(t)u_5(0;t)$ and
$u^\prime(0;t)=c_1(t)u_1^\prime(0;t)+c_5(t)u_5^\prime(0;t)$ as the
fundamental data for the problem. Choosing $c_1(t), c_5(t)$ so
that the inverse Mellin transforms of these terms exist on a strip
of the complex plane implies (by the work done above) that the
inverses of $u$ and $u^\prime$ also exist on the space-like
hypersurface $z=1/2$. Then verifying that the restriction on $t$
implies finiteness of the basis solutions $u_2,u_6$ at $z=1$, we
can finally conclude that the inverse Mellin transforms of $P$ and
$P^\prime$ exist at the future null cone $z=1$.

The thrust of this argument is that the `scattering coefficients'
$c_i(t)$ {\em do not} affect resummability of the solution, and so
that finiteness of the basis solutions $u_2,u_6$ at $z=1$ is a
necessary and also {\em sufficient} condition for finiteness of
the field $\phi$ at the future null cone.

Apart from actually verifying these assertions rigorously, the
last step in our argument involves showing that the scenario
summarized in the two preceding paragraphs carries over to Vaidya
space-time. The crucial step is to consider the $t-$dependence of
pairs of fundamental solutions of the ODE at the two crucial
singular points: $(p_1^N, p_2^N)$ at the past null cone and
$(p_1^C,p_2^C)$ at the Cauchy horizon (future null cone). We
consider first solutions at $x=0$.

The standard form of (\ref{melltreq}) at $x=0$ is
\[ x^2P^{\prime\prime}+xq(x;t)P^\prime+r(x;t)P=0,\]
where
\bes q(x;t)&=&\frac{2t-2tx+(1+2t)\lambda x^2}{2-x+\lambda x^2},\\
r(x;t)&=&-\frac{(t^2-t-l(l+1)-t^2\lambda x)x}{2-x+\lambda
x^2}.\ees Note that these are analytic at $x=0$ and can be written
in the form \[ q(x;t)=\sum_{k=0}^\infty q_k(t)x^k,\qquad
r(x;t)=\sum_{k=0}^\infty r_k(t)x^k,\] where convergence of the
series is guaranteed for $x\in[0,x_c)$. The indicial equation is
\[ I(\nu):=\nu(\nu-1)+q_0\nu+r_0=\nu(\nu-1+t)=0,\]
with solutions $\nu=0,1-t\not\in\mathbb{Z}$. Thus we have the
linearly independent solutions \bes
p_1^N(x;t)&=&\sum_{k=0}^\infty a_k(t)x^k,\quad a_0=1,\\
p_2^N(x;t)&=&|x|^{1-t}\sum_{k=0}^\infty b_k(t)x^k,\quad b_0=1.\ees
In each case, the radius of convergence of the series is at least
$x_c$. Thus this representation of a fundamental set of solutions
is valid for all $x\in[0,x_c)$. The recurrence relations for the
$a_k(t)$ are
\[
a_k(t)=-\frac{1}{I(k)}\sum_{j=0}^{k-1}(jq_{k-j}(t)+r_{k-j}(t))a_j(t),\quad
k\geq 1,\] and the recurrence relations for the $b_k$ are
\[
b_k(t)=-\frac{1}{I(k+1-t)}\sum_{j=0}^{k-1}((j+1-t)q_{k-j}(t)+r_{k-j}(t))b_j(t),\quad
k\geq 1.\] We wish to determine the $t-$dependence of these
coefficients. Noting that the coefficients $q_k(t)$ and $r_k(t)$
are respectively linear and quadratic in $t$ and that $I(\nu)$ is
linear in $t$, it is straightforward to prove that
\[ a_k(t) = \frac{\cp_{2k}(t)}{\cp_k(t)},\quad
b_k=\frac{\cp_{2k}(t)}{\cp_k(t)},\] where $\cp_k(t)$ is used to
represent an arbitrary polynomial of degree $k$ which may be
different in different formulae and within a single formula. Then
division in the ring of polynomials yields
\[ a_k=\cp_k(t)+\frac{\cp_{k-1}(t)}{\cp_k(t)},\]
with a similar result for $b_k$. From here it is clear that we can
write more explicitly \bes a_k
&=&\cp_k(t)+\frac{\cp_{k-1}(t)}{t(t+1)\cdots(t+k-1)}=\cp_k(t)+\sum_{j=0}^{k-1}\frac{a_k^j}{t-j},\\
b_k&=&\cp_k(t)+\frac{\cp_{k-1}(t)}{(2-t)(3-t)\cdots(k+1-t)}=\cp_k(t)+\sum_{j=0}^{k-1}\frac{b_k^j}{j+1-t},\ees
where the $a_k^j$ and $b_k^j$ are independent of $t$.

Before proceeding, we note another property of the Mellin
transform that will allow us to deal with the inversion of terms
arising from the presence of the $\cp_k(t)$ in the coefficients
$a_k(t)$.

\begin{lemma}
If \[ G(t)={\cal{M}}[\gamma(r)](r\to t),\] then, provided the
relevant integrals converge,
\[s^nG(t)={\cal{M}}[D^n\gamma(r)](r\to t)\]
where the differential operator is defined by
$D=-r\frac{\partial}{\partial r}$.
\end{lemma}
\noindent{\bf Proof:} This follows by induction and by using Lemma
6.1.

There is an immediate corollary:

\begin{corollary}
If\[ G(t)={\cal{M}}[\gamma(r)](r\to t),\] and
$p(t)=p_0+p_1t+\cdots+p_nt^n$ is a polynomial in $t$, then
\[ p(t)G(t)={\cal{M}}[(p_0+p_1D+\cdots+p_nD^n)\gamma(r)](r\to
t).\]
\end{corollary}

Note that the preceding analysis applies equally well to the case
$\lambda=0$, i.e.\ to the wave equation in Minkowski space-time.
Thus the solutions $u_1,u_5$ can equally well be written as
combinations of $p_1^N,p_2^N$. We emphasize this as we now have
solutions in Minkowski space-time that involve infinite series
rather than polynomials.

The solutions $p_i^N, i=1,2$ `scatter' to a naturally arising
fundamental set of linearly independent solutions $p_1^C, p_2^C$
defined at $x=x_c$ (where $x_c=2$ in Minkowski space-time). Indeed
by writing (\ref{melltreq}) in standard form at $x=x_c$ and
determining the roots of the indicial equation thereat, we can
write explicitly
\be p_1^C(x;t)&=&\sum_{n=0}^\infty A_k(t)(x-x_c)^n,\\
p_2^C(x;t)&=&|x-x_c|^{\nu_2}\sum_{n=0}^\infty
B_k(t)(x-x_c)^n,\label{p2solch}\ee where
\[
\nu_2=\frac{(1-8\lambda+\sqrt{1-8\lambda})(1-t)}{2-16\lambda}=1-t+O(\lambda)\]
and both series converge in some neighbourhood of $x=x_c$. We note
that $\mathrm{Re}(\nu_2)>\mathrm{Re}(1-t)$, and so the earlier
restriction $\mathrm{Re}(t)<0$ implies that
$\mathrm{Re}(\nu_2)>1$, and so both solutions $p_i^C, i=1,2$ are
finite at the Cauchy horizon. (This applies quite generally in
self-similar collapse to a naked singularity when the dominant
energy condition holds: see \cite{bandt}.) As these solutions can
be written as {\em different} $\mathbb{C}-$linear combinations of
$p_i^N, i=1,2$, this implies that the series representations of
$p_i^N, i=1,2$ {\em must both converge at} $x=x_c$. This is of
crucial importance to our argument, as we can now determine the
field $\phi(x,r)$ at the future null cone $x=x_c$ by carrying out
(i.e.\ checking convergence of) the inverse Mellin transform of
the sum of the convergent series \be
P(x_c;t)=a(t)p_1^N(x_c;t)+p_2^N(x_c;t).\label{tersol}\ee Here
$a(t),b(t)$ are functions whose inverse Mellin transforms
$\alpha(r),\beta(r)$ exist for inversion contours lying in some
strip of the complex plane. $\alpha,\beta$ constitute Cauchy data
for $\phi$, and satisfy unspecified differentiability and
integrability conditions.

As we have seen, this inverse Mellin transform will involve an
infinite series of terms of the form
\[ \alpha_k(r) =
\cp_k(D)\alpha(r)+\sum_{c_i}k_ir^{c_i}\int_0^r\frac{\alpha(y)}{y^{c_i-1}}\,dy,\]
where $D$ is the differential operator introduced above. There
will be similar terms arising from the contribution by $\beta(r)$,
the inverse Mellin transform of $b(t)$. Existence of these
individual terms can be guaranteed by imposing differentiability
and integrability conditions on the initial data functions
$\alpha(r),\beta(r)$. (Assuming differentiability of arbitrarily
high order should not be a significant constraint here as
linearity would allow us to work in distributions or to use a
density argument to generalise from analytic functions to more
interesting spaces. We also note that the sign of the $c_i$ in the
$\alpha_k$ above will not be of particular relevance, as for
either case of this sign, either the multiplicative pre-factor or
the divisor in the integrand will have a mollifying effect.) Hence
our problem boils down to deducing convergence of series of the
form $\sum_{k=0}^\infty \alpha_k(r)$. Here is another crucial
point: our analysis of the problem in Minkowski space-time from a
slightly different point of view {\em guarantees} that such series
must converge in the case $\lambda=0$.

Now the coefficients of these series are generated by the
coefficients of the series in (\ref{tersol}). The only difference
between Minkowski space-time and Vaidya space-time is the value of
$\lambda$ ($\lambda=0$ and $\lambda\in(0,1/8)$ respectively). For
$\lambda=0$, the coefficients of the series of complex numbers
(\ref{tersol}) guarantees convergence in $\mathbb{C}$: these
coefficients generate coefficients of an infinite series $\sum
(c_k\alpha_k(r)+d_k\beta_k(r))$ in a certain unspecified function
space - call it $\mathcal{F}$ - that, as we have argued, guarantee
convergence in this space. {\em Convergence of this series depends
only on the coefficients inherited from (\ref{tersol}).} To
conclude our argument, we maintain that this pattern is repeated
when $\lambda>0$: the coefficients of the $\mathbb{C}-$convergent
series (\ref{tersol}) generate coefficients of series of functions
in a certain unspecified function space that must then also
converge in this function space.

We conclude this section by summarising the argument. Restricting
the values of $t$ to allow only a finite flux at the regular axis
and at the past null cone is sufficient in the present case to
guarantee finiteness of the Mellin transform $P(x;t)$ of the field
$\phi$ at the future null cone. To determine if $\phi$ itself is
finite at the future null cone, we must calculate the inverse
Mellin transform of $P$. The analytic form of the line element of
Vaidya space-time allows us to determine the general form of this
inverse: it involves an infinite series of finite sums of
derivatives and integrals of the initial data for $\phi$. As these
finite sums converge, convergence of the full series of functions
in $\mathcal{F}$ depends only on the constant coefficients of the
series: these coefficients are generated by the coefficients in
the convergent $\mathbb{C}-$series representation of $P(x;t)$ at
$x=x_c$. In Minkwoski space-time, the coefficients
$\{p_k(0)\}_{k=0}^\infty$ of the series $P(x_c;t)$ of complex
numbers produces a convergent series representation for
$\phi|_{x=x_c}$ in $\mathcal{F}$. We claim that in Vaidya
space-time, the same will happen: the coefficients
$\{p_k(\lambda)\}_{k=0}^\infty$ of the series $P(x_c;t)$ of
complex numbers produces a convergent series representation for
$\phi|_{x=x_c}$ in $\mathcal{F}$.

This last paragraph constitutes our argument that in order to
determine finiteness of the field at the Cauchy horizon, it is
sufficient to check finiteness of the modes thereat. There clearly
remains a good deal to prove, but the analysis above suggests a
way of doing this: the main thing that requires checking is the
convergence of the finite sums $\alpha_k(r)$ and of the overall
series $\sum\alpha_k$. Finally we note that although this argument
has been presented for only the wave equation, it should
generalise to any set of linear equations, as for example the
perturbation equations considered in this paper.

\section{Conclusions.}

We have considered metric and matter perturbations of all angular
modes falling on the Cauchy horizon formed by the naked
singularity arising from the collapse of a self-similar null dust.
There is no class of perturbation which satisfies the initial
conditions and gives a divergent flux on the Cauchy horizon, thus
there is no ``blue sheet'' instability as is seen in, for example,
the Reissner-Nordstr\"om naked singularity. (We have shown
rigorously that the modes are finite at the Cauchy horizon, and
given a plausibility argument that the full perturbation itself,
found by resumming over the modes, is finite.) Interestingly, the
second future similarity horizon of the self-similar null dust
space-time \emph{is} unstable for perturbations with multipole
modes $l \geq 2$.

The question of uniqueness of the perturbation to the future of
the Cauchy horizon has not been addressed, but this should not
affect the divergence encountered at the SFSH. If we consider the
wave equation as studied in Section 6 as a paradigm, we note that
the solution (\ref{p2solch}) and its derivative vanish at the
Cauchy horizon for the allowed range of $t$ ($\mathrm{Re}(t)<0$).
Thus an arbitrary additional amount of $p_2^C$ could be added to
the solution for $x>x_c$ while preserving continuity and
differentiability. However in Minkowski space-time a selection
procedure tells us the correct addition to make, based on
reflection through the (still) regular axis. It is not clear that
one can apply a similar argument in Vaidya space-time, where one
encounters a singularity at $r=0$ to the future of the future null
cone (Cauchy horizon): indeed this is the central problem caused
by a naked singularity, and the very issue that the Cosmic Censor
seeks to render irrelevant. Whether establishing uniqueness is
possible or not, some amount of both independent solutions $p_i^C,
i=1,2$ will persist in $x>x_c$, leading to divergence at $x=x_e$.

It is worrying from the point of view of the Cosmic Censorship
Hypothesis that this naked singularity persists. However, this
worry is tempered by the fact that an instability is encountered
at the SFSH of the space-time; the naked singularity only survives
for a finite time. It is then of interest to consider the
life-time of this naked singularity and what effects it might have
on the space-time. It is also of interest to know how generic this
short-lived stability of the naked singularity is. In the
terminology of Carr and Gundlach \cite{CarrGund}, this second
future similarity horizon is a ``splash'', whereas the other
similarity horizons are ``fans''. We speculate that finiteness of
the flux of perturbations on a fan-type similarity horizon, and
divergence on a splash-type similarity horizon, may be a general
feature of self-similar spherically symmetric space-times; the
issue is currently being studied.

The results here also have some bearing on the issue of stability
in critical collapse. In perfect fluid collapse, the critical
space-times are continuously self-similar, i.e. admit a homothetic
killing vector field (see e.g.\ \cite{carstenreview}). By
definition, the critical space-times posses a single (spherically
symmetric) unstable mode for perturbations injected along the
homothetic surfaces $x=$ constant and followed up to the
singularity. In our variables, this corresponds to a mode of the
form $r^sQ(x)$, with the limit $r\to 0$ taken along $x=$ constant,
and with $x<x_c$. Then $Re(s)<0$ corresponds to instability. In
the present paper, the limiting behaviour of the perturbation has
been studied in the approach to the first and second future
similarity horizons in Vaidya space-time. Carrying out a similar
calculation for the critical fluid space-times may reveal very
different stability properties to that shown by critical collapse
studies.

\vspace{0.5in}

\section*{Acknowledgement}
The authors would like to thank Patrick Brady, Tomohiro Harada and
T.P. Singh for helpful conversations. This research is supported
by Enterprise Ireland grant SC/2001/199.

\appendix
\renewcommand{\theequation}{\Alph{section}.\arabic{equation}}
\section{Perturbation equations in terms of gauge invariants.}

We give here the full set of perturbation equations for the gauge
invariants defined in (\ref{gis}). Note not all equations apply
for $l=0,1$ modes.
\begin{subequations} \label{minkpert}
\begin{eqnarray}
2\w{v}^C(k_{AB|C}-k_{CA|B}-k_{CB|A}+2\w{g}_{AB}k_{CD}^{\phantom{CD}|D})-\left(\f{l(l+1)}{r^2}+\w{G}_C^{\phantom{C}C}+\w{G}_a^{\phantom{a}a}+2\w{\mathcal{R}}\right)k_{AB}&&\nonumber\\
+\w{g}_{AB}\left(\f{l(l+1)}{r^2}+\f{1}{2}(\w{G}_C^{\phantom{C}C}+\w{G}_a^{\phantom{a}a})+\w{\mathcal{R}}\right)k_D^{\phantom{D}D}-2\w{g}_{AB}\w{v}^C
k_{D\phantom{D}|C}^{\phantom{D}D}+\w{g}_{AB}(2 \w{v}^{C|D}+4\w{v}^C \w{v}^D-\w{G}^{CD})k_{CD}&&\nonumber\\
-\w{g}_{AB}\left(2k,_C^{\phantom{,C}|C}+6\w{v}^C k,_C-\f{(l-1)(l+2)}{r^2}k\right)+2(\w{v}_Ak,_B+\w{v}_Bk,_A+k,_{A|B})=-16 \pi T_{AB},&(l\geq0)\qquad\\
-(k,_C^{\phantom{C}|C}+2\w{v}^Ck,_C+\w{G}_a^{\phantom{a}a}k)+\left(k_{CD}^{\phantom{CD}|C|D}+2\w{v}^Ck_{CD}^{\phantom{CD}|D}+2(\w{v}^{C|D}+\w{v}^C \w{v}^D)k_{CD}\right)&&\nonumber\\
-\left(k_{C\phantom{C}|D}^{\phantom{C}C\phantom{|D}|D}+\w{\mathcal{R}} k_C^{\phantom{C}C}-\f{l(l+1)}{2r^2}k_C^{\phantom{C}C}\right)=-16 \pi T^1, &(l\geq0)\qquad\\
k,_A-k_{AC}^{\phantom{AC}|C}+k_{C\phantom{C}|A}^{\phantom{C}C}-\w{v}_A k_C^{\phantom{C}C}=-16\pi T_A,&(l\geq1) \qquad\\
k_A^{\phantom{A}A}=-16\pi T^2.&(l\geq2)\qquad \label{tracek}
\end{eqnarray}
\end{subequations}

Here $\tilde{v}^A=\tfrac{r^{|A}}{r}$, and $\w{G}_{\mu\nu}$ is the
Einstein tensor of the background space-time. $\w{\mathcal{R}}$ is
the Gaussian curvature of $M^2$, the manifold spanned by the time
and radial coordinates, and thus equals half the Ricci scalar of
$M^2$.

\section{First order system of perturbation equations.}

The perturbation equations give rise to the first order system
$Y'=M(x)Y$ where $Y=(A,D,K,G)^T$ and the coefficients of $M$ are:

\begin{small}
\be
\begin{array}{ll}
\multicolumn{2}{l}{M_{11}=\frac{4 - 4\,x + 2\,\left( 1 + \lambda
\right) \,x^2 - 3\,\lambda\,x^3 + \lambda^2\,x^4 + 2\,s^2\,\left(
2 - x + \lambda\,x^2 \right) +
    s\,\left( -4 + 2\,x - \lambda\,x^3 + \lambda^2\,x^4 \right) }{x\,\left( 2\,s + \lambda\,x^2 \right) \,\left( 2 - x + \lambda\,x^2 \right)
    }}\\
\multicolumn{2}{l}{M_{12}= \frac{4\,\left( -1 + x \right)  -
2\,s^2\,x\,{\left( -1 + x\,\lambda  \right) }^2 +
    l\,\left( 2 + x\,\left( -1 + x\,\lambda  \right)  \right) \,
     \left( 2 + x\,\left( -1 + x\,\lambda  \right) \,\left( 2 + x\,\left( -1 + x\,\lambda  \right)  \right)  \right)  +
    l^2\,\left( 2 + x\,\left( -1 + x\,\lambda  \right)  \right) \,
     \left( 2 + x\,\left( -1 + x\,\lambda  \right) \,\left( 2 + x\,\left( -1 + x\,\lambda  \right)  \right)  \right) }{x^2\,
    \left( -1 + x\,\lambda  \right) \,\left( 2\,s + x^2\,\lambda  \right) \,\left( 2 + x\,\left( -1 + x\,\lambda  \right)  \right) }}  \\
\multicolumn{2}{l}{\qquad\quad\frac{-\left( s\,\left( -1 +
x\,\lambda \right) \,\left( 4 +
         x\,\left( -2 + x\,\left( 2 + \lambda \,\left( -2 + x\,\left( -5 + 3\,x\,\lambda  \right)  \right)  \right)  \right)
         \right)  \right)  + x\,\left( 8\,\lambda  +
       x\,\left( -2 + \lambda \,\left( -12 + x\,\left( 9 + \lambda \,\left( 8 + 7\,x\,\left( -2 + x\,\lambda  \right)  \right)
                \right)  \right)  \right)  \right) }{x^2\,\left( -1 + x\,\lambda  \right) \,\left( 2\,s + x^2\,\lambda  \right) \,
    \left( 2 + x\,\left( -1 + x\,\lambda  \right)  \right) }}  \\
\multicolumn{2}{l}{M_{13}= \frac{\left( 2 + \left( -2 + l + l^2
\right) \,x \right) \,\left( 2 + \left( -2 + x \right) \,x \right)
-
    x^2\,\left( -6 + \left( 9 + 2\,l\,\left( 1 + l \right) \,\left( -1 + x \right)  - 4\,x \right) \,x \right) \,\lambda  +
    x^4\,\left( 3 + \left( -2 + l + l^2 \right) \,x \right) \,\lambda
    ^2    }{ x^2\,
    \left( 2\,s + x^2\,\lambda  \right) \,\left( 2 + x\,\left( -1 + x\,\lambda  \right)  \right) } }  \\
\multicolumn{2}{l}{  \qquad\quad   \frac{-2\,s^2\,\left( 2 +
x\,\left( -1 + x\,\lambda \right)  \right)  +
    s\,x\,\left( 2 - x\,\left( 2 + \lambda \,\left( 4 + x\,\left( -5 + 3\,x\,\lambda  \right)  \right)  \right)  \right) }{x^2\,
    \left( 2\,s + x^2\,\lambda  \right) \,\left( 2 + x\,\left( -1 + x\,\lambda  \right)  \right)}}  \\
\multicolumn{2}{l}{    M_{14}= \frac{2\,\lambda \,\left( -4 +
2\,s\,\left( 2 + x\,\left( -1 + x\,\lambda  \right)  \right)  -
      x\,\left( -2 + x\,\lambda  \right) \,\left( 2 + x\,\left( -1 + x\,\lambda  \right)  \right)  \right) }{x\,
    \left( 2\,s + x^2\,\lambda  \right) \,\left( 2 + x\,\left( -1 + x\,\lambda  \right)  \right) } } \\
\multicolumn{2}{l}{    M_{21}= \frac{2\,x\,\left( -1 + x\,\lambda
\right) }
  {\left( 2\,s + x^2\,\lambda  \right) \,\left( 2 + x\,\left( -1 + x\,\lambda  \right)  \right) }}  \\
\multicolumn{2}{l}{    M_{22}= \frac{2 + l\,\left( 1 + l \right)
\,\left( -2 + x \right)  -
    \left( \left( -1 + s \right) \,\left( -6 + 4\,s - x \right)  + 2\,l\,\left( -1 + x \right)  +
       2\,l^2\,\left( -1 + x \right)  \right) \,x\,\lambda}{\left( -1 + x\,\lambda  \right) \,
    \left( 2\,s + x^2\,\lambda  \right) \,\left( 2 + x\,\left( -1 + x\,\lambda  \right)  \right)
    }}  \\
\multicolumn{2}{l}   {  \qquad\quad     + \frac{    x^2\,\left( 6
+ 2\,s\,\left( -3 + s - x \right)  + \left( 3 + l + l^2 \right)
\,x \right) \,{\lambda }^2 +
    \left( -2 + s \right) \,x^4\,{\lambda }^3 + 2\,s\,\left( -2 + s + 2\,\lambda  \right) }{\left( -1 + x\,\lambda  \right) \,
    \left( 2\,s + x^2\,\lambda  \right) \,\left( 2 + x\,\left( -1 + x\,\lambda  \right)  \right) }}  \\
    M_{23}= - \frac{\left( -2 + 2\,s - \left( -2 + l + l^2 \right) \,x \right) \,\left( -1 + x\,\lambda  \right) }
    {\left( 2\,s + x^2\,\lambda  \right) \,\left( 2 + x\,\left( -1 + x\,\lambda  \right)  \right) }
   & \hspace{1in} M_{24}= \frac{-4\,x\,\lambda \,\left( -1 + x\,\lambda  \right) }
  {\left( 2\,s + x^2\,\lambda  \right) \,\left( 2 + x\,\left( -1 + x\,\lambda  \right)  \right) }  \\
    M_{31}= \frac{-2}{2\,s + x^2\,\lambda } &
   \hspace{1in} M_{32}= \frac{2 - 2\,s + l\,\left( 1 + l \right) \,\left( -2 + x \right)  -
    x\,\left( 4 - 2\,s + l\,\left( 1 + l \right) \,x \right) \,\lambda }{x\,\left( -1 + x\,\lambda  \right) \,
    \left( 2\,s + x^2\,\lambda  \right) }  \\
    M_{33}= \frac{-2 + 2\,s^2 + s\,x^2\,\lambda  - x\,\left( -2 + l + l^2 + x\,\lambda  \right) }{2\,s\,x + x^3\,\lambda }
    &
   \hspace{1in} M_{34}= \frac{4\,\lambda }{2\,s + x^2\,\lambda }  \\
    M_{41}= 0  &
   \hspace{1in} M_{42}= \frac{1}{-x + x^2\,\lambda }  \\
    M_{43}= 0 &
   \hspace{1in} M_{44}= \frac{s}{x}  \\
\end{array}
\ee
\end{small}

\section{Methods for systems of ordinary differential equations with singular points.}

For a first order system $Y'=M(x)Y$ we define $p$ as the least
number such that the system can be written near $x=0$ as \be
Y'=\frac{1}{x^p}\left(J+\sum_{m=1}^{\infty}A_mx^m\right)Y,
 \label{singzero} \ee and near $x=\infty$ as \be
Y'=-x^{p-2}\left(J+\sum_{m=1}^{\infty}A_mx^{-m}\right)Y.
 \label{singinf}   \ee with $J \neq 0$ and constant.

We classify singular points and describe the solution as:

\begin{description}
\item[$\mathbf{p=0}$, Regular point.] The solutions of a system of
differential equations are at least as regular as the coefficients
of the system of differential equations. For the purposes of this
paper, it is enough to know that solutions near a regular point
are themselves regular, as none of the surfaces of interest are
regular points.

\item[$\mathbf{p=1}$, Regular singular point.] Also known as a
simple singularity or a singularity of the first kind. Here we
distinguish solutions depending on whether the eigenvalues of $J$
given above differ by an integer or not. If they do not, we apply
Theorem \ref{theorem:thone} immediately. If they do, we reduce
those eigenvalues individually until they are equal using Theorem
\ref{theorem:thtwo}, and then apply Theorem \ref{theorem:thone}.
(see \cite{CoddLev}).

\begin{theorem} \label{theorem:thone} In the system (\ref{singzero}), if $J$ has
eigenvalues which do not differ by positive integers, then, in a
disc around $x=0$ not containing another singular point,
(\ref{singzero}) has a fundamental matrix $\Phi$ of the form \be
\Phi=P x^J, \qquad \textrm{where} \quad P(x)=\sum_{m=0}^{\infty}
x^m P_m, \qquad P_0=I. \ee
\end{theorem}

\begin{theorem} \label{theorem:thtwo} Let the distinct eigenvalues of $J$ in
(\ref{singzero}) be (disregarding multiplicity) $\mu_1, \ldots
\mu_k$, ($k\leq N$, where $N$ is the order of the system). There
is a matrix $V(x)$ such that $Y=VQ$ transforms (\ref{singzero})
into \be Q'=\frac{1}{x}\left(\wh{J}+\sum_{m=1}^{\infty}\wh{A}_m
x^m\right)Q, \ee where $\wh{J}$ has eigenvalues $\mu_1-1,
\mu_2,\ldots ,\mu_k$. $V$ is given by a diagonal matrix with
entries 1, except where the eigenvalue(s) to be reduced occurs,
where there is an $x$.
\end{theorem}

\item[$\mathbf{p\geq 2}$, Irregular singular point.] Also known as
a non-simple singularity or a singularity of the second kind. Here
we distinguish solutions depending on whether you can diagonalize
$J$ given in (\ref{singinf}). If $J$ has distinct eigenvalues,
then $J$ is diagonalizable and we apply Theorem
\ref{theorem:ththree}. If $J$ has multiple eigenvalues and $J$ can
only be reduced to Jordan normal form, then we apply Theorem
\ref{theorem:thfour} to remove off-diagonal terms (see
\cite{Eastham}). When the eigenvalues are repeated zeroes, this
has the effect of reducing the order of the singularity, as
happens in this paper at the threshold (see part V).\\ There is a
class of problems in between: sometimes a matrix has multiple
eigenvalues and yet can still be diagonalized. In this case, there
is a straightforward Theorem given by \cite{Eastham} if $A_1=0$
(as is sometimes the case when a high order equation is written as
a first order system). If not there is a very cumbersome solution
given by \cite{CoddLev}.

\begin{theorem} \label{theorem:ththree} Let $J$ have distinct eigenvalues $\mu_1, \ldots
,\mu_N$. Then (\ref{singzero}) has a fundamental matrix \be \Phi
=P x^R e^H, \qquad \textrm{where} \quad P(x)=\sum_{m=0}^{\infty}
x^m P_m, \qquad P_0=I, \ee $R$ is a diagonal matrix of complex
constants, and $H$ is a matrix polynomial ($r=p-2$) \be
H=\f{x^{r+1}}{r+1}H_0+\f{x^r}{r}H_1+\ldots +xH_r,\qquad
H_i=\textrm{diag}\left(\mu_1^{(i)},\ldots, \mu_N^{(i)}\right),
\quad \mu_j^{(0)}=\mu_j. \ee
\end{theorem}

\begin{theorem} \label{theorem:thfour} For brevity's sake, we give the Theorem only for
$p=2$ as this is the case that arises in this paper. We transform
$J$ to it's Jordan normal form $\wh{J}$, and write the blocks of
$\wh{J}$ as $\mu I+\rho E$, where $E$ is the matrix with 1's along
it's super-diagonal and zeroes elsewhere. For each block of
$\wh{J}$, define the matrices \be D=\textrm{diag}(1,\rho
x,\ldots,(\rho x)^{N-1}), \qquad
B=\left(\begin{array}{ccccc} 1 & 1 & 1/2! & \ldots & 1/(N-1)! \\
                               0 & 1 & 1 & \ldots & 1/(N-2)! \\
                                 &   & \ddots  & \ddots &  \vdots
                                 \\ &&&& 1
\end{array} \right),
\ee where $N$ is the order of the system. Then the transformation
$Y=D^{-1} B W$ gives the system \be W'=\left[\mu I+D'
D^{-1}+B^{-1} D \left(\sum_{m=1}^{\infty}A_m x^{-m} \right) D^{-1}
B \right] W, \ee and the leading order coefficient matrix has had
its off-diagonal terms removed.
\end{theorem}

\end{description}

\end{document}